\documentclass{article}

\usepackage{preambles/PRIMEarxiv}

\usepackage[utf8]{inputenc} % allow utf-8 input
\usepackage[T1]{fontenc}    % use 8-bit T1 fonts
\usepackage{hyperref}       % hyperlinks
\usepackage{url}            % simple URL typesetting
\usepackage{booktabs}       % professional-quality tables
\usepackage{amsfonts}       % blackboard math symbols
\usepackage{nicefrac}       % compact symbols for 1/2, etc.
\usepackage{microtype}      % microtypography
\usepackage{lipsum}
\usepackage{fancyhdr}       % header
\usepackage{graphicx}       % graphics
\graphicspath{{media/}}     % organize your images and other figures under media/ folder

\usepackage{tikz}
\usetikzlibrary{shapes, arrows.meta, positioning, fit, trees, decorations, spy, calc, backgrounds}
\usepackage{enumitem}
\setlist[enumerate]{font=\sffamily}
\usepackage{soul}
\usepackage[square,numbers]{natbib}
\usepackage{threeparttable}
\usepackage{tcolorbox}
\usepackage{adjustbox}
\usepackage{pgfplots}
\pgfplotsset{compat=1.17}
\pgfplotsset{
  colormap={divRedWhiteBlue}{
      color(0cm)=(red);
      color(0.5cm)=(white);
      color(1cm)=(blue)
  }
}
\usepackage{amsmath}
\usepackage{amssymb}
\usepackage{caption}
\usepackage{bigints}
\usepackage{lmodern}
\usepackage{xcolor}
\usepackage{dsfont}
\usepackage{bbm}

\hypersetup{
    colorlinks = {true},
    citecolor = {blue},
    menubordercolor = {red},
    citebordercolor = {white},
    linkbordercolor = {red},
    urlcolor = {black}
}

\newcommand{\tokenizertkn}[2]{\adjustbox{valign=M, margin=0pt}{\colorbox{#1}{\strut#2}}}
\newcommand{\token}[1][]{\ensuremath{\textsf{d}_{#1}}}
\newcommand{\tokentotal}{\ensuremath{\textsf{d}_{\max}}}
\newcommand{\plotElement}[5]{%
    \begin{scope}[shift={(#1, #2)}]
        \node[draw=black, thick, minimum width=1cm, minimum height=1.5cm, inner sep=0] (rect) {};
        \foreach \i in {0,1,2,3} {
            \draw[fill=#4!10, draw=#4!80]
                ($(rect.north west)+(0.1, -0.1-\i*0.35)$) rectangle ++(0.8, -0.25);
        }
        \node[#5=0mm of rect] {#3};
    \end{scope}
}

\title{Decoding Consumer Preferences Using Attention-Based Language Models
}

\author{
  Joshua Foster, Fredrik \O{}degaard \\
  Ivey Business School \\
  Western University \\
  London, Ontario, Canada\\
  \texttt{\{jfoster, fodegaard\}@ivey.ca} \\
}

\begin{document}
\maketitle

\begin{abstract}
This paper proposes a new demand estimation method using attention-based language models. An encoder-only language model is trained in a two-stage process to analyze the natural language descriptions of used cars from a large US-based online auction marketplace. The approach enables semi-nonparametrically estimation for the demand primitives of a structural model representing the private valuations and market size for each vehicle listing. In the first stage, the language model is fine-tuned to encode the target auction outcomes using the natural language vehicle descriptions. In the second stage, the trained language model's encodings are projected into the parameter space of the structural model. The model's capability to conduct counterfactual analyses within the trained market space is validated using a subsample of withheld auction data, which includes a set of unique ``zero shot'' instances. 
\end{abstract}

% keywords can be removed

\section{Introduction} 

Identifying the determinants of consumer value is a fundamental challenge of understanding market demand. For many products, commercial success hinges on how well their features reflect the underlying consumer preferences in the market, and thus firms have a strong economic incentive to determine how various combinations of features produce value. One common approach to estimating the strength of consumer preferences is to convert raw, unstructured feature data into a structured dataset that could plausibly delineate the sources of all value. For instance, in a used car market one might collect data on vehicles' make, model, year, and miles, while for theater shows one might collect data on the type of show, cast, venue seat, and day-of-week. Using these data, prices can be modeled as a function of the overall value, as estimated by summing the {\it hedonic} values of the features in the product \citep{rosen1974hedonic}. 

\begin{figure}
\centering 
\resizebox{\textwidth}{!}{
  \includegraphics[]{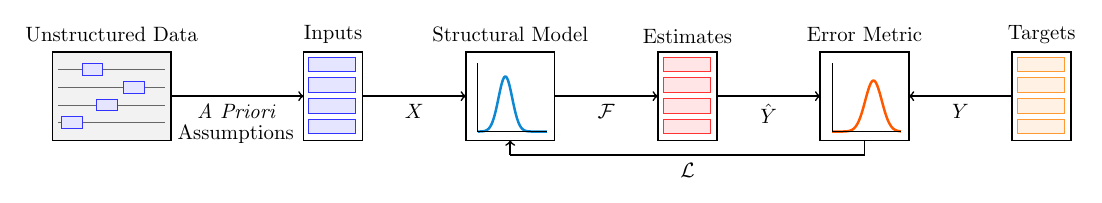}
}
\caption{Standard Estimation Model}
\label{fig:standard-model}
\end{figure}

As a generic illustration, consider a standard approach presented in Figure \ref{fig:standard-model}. First, a set of features $X$ is defined from an unstructured data source (e.g., vehicle description in a used car market) to serve as inputs into a structural estimation model $\mathcal{F}$. The structural model produces estimates $\hat{Y}$ on targets $Y$ (e.g., sales price of used cars), and an error metric $\mathcal{L}$ is used to optimize the parameters of the structural model (e.g., beta-coefficients of OLS regression). 

Although simple and intuitive, one drawback to this approach is that it often involves making {\it a priori} assumptions on the sources and structure of consumer valuations. Without sufficient knowledge of how to code the true sources of value, one must simply rely (perhaps unscientifically) on intuition for this task \citep{ludwig2024machine}. And, even with perfect knowledge of the feature set, feature-based preferences can be diverse, dynamic, and interdependent -- potentially requiring complex identification methods in order to estimate the true hedonic values of consumers \citep{cropper1988choice}. Furthermore, any resulting insights will be limited to the defined attributes and specific market dynamics, as it is not immediate how, for instance, a standard regression-based inference might apply to an out-of-sample good, or a set of previously unobserved attributes, or under different market mechanisms.  
 
Motivated by the above limitations and the recent development and applications of natural language processing, we offer a new approach to modeling the sources of consumer valuations. Specifically, we feed a detailed text description of a good within a specific market environment to an attention-based language model, which converts the text into {\it embedding} vectors of a high-dimensional, real-valued space that can numerically represent the good's underlying demand and market information. We label this space as the {\it Value Embedding Space}, and each specific embedding vector, corresponding to the description of a particular good, as a {\it Demand Embedding Vector} (DEV). The benefit of this language-model-based approach is that it can process and interpret intricate language patterns, contextual information, and nuanced connotations of product features, and thus generate more complex representations of consumer preferences than traditional methods. 

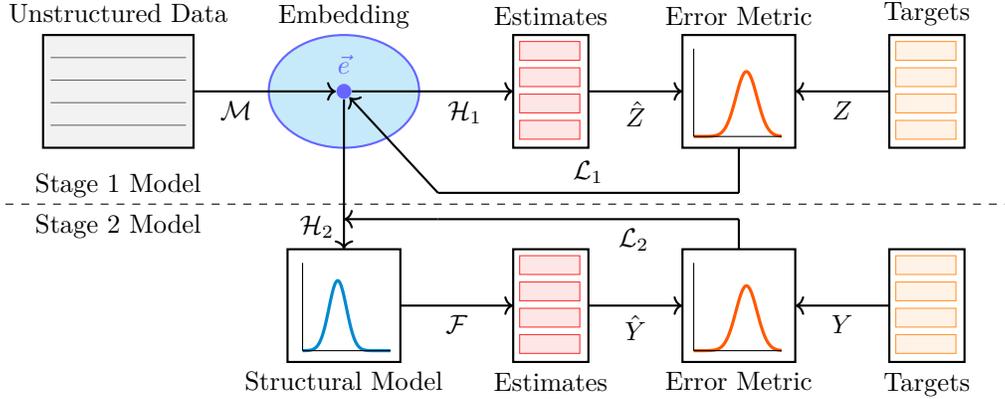
\begin{figure}
\centering 
\begin{tikzpicture}
  %%% Stage 1 %%% 

  % Draw a piece of paper with broken up horizontal lines
  \node at (-4, -2.2) {Unstructured Data};
  \draw[thick, fill=gray!10, draw=black] (-5,-4) rectangle (-3,-2.5);
  \foreach \y in {-3.7, -3.4, -3.1, -2.8} {
    \draw[-, draw=black!60] (-4.9, \y) -- (-3.1, \y);
  }
  
  % Draw an oval-shaped set space
  \node at (-1, -2.25) {Embedding};
  \draw[thick, fill=cyan!20, draw=blue!60] (-1,-3.25) ellipse (1cm and 0.75cm);
  \node[fill=blue!60, circle, inner sep=2pt, label=above:{\textcolor{blue!60}{$\vec{e}$}}] (e) at (-1.0,-3.25) {};

  % Estimates 
  \plotElement{1.75}{-3.25}{Estimates}{red}{above}

  % Error Metric 
  \node at (4.25, -2.25) {Error Metric};
  \draw[thick] (5, -4) rectangle (3.5, -2.5);
  \begin{axis}[
    name=plot1,
    at={(3.65cm, -3.85cm)},
    anchor=south west,
    no markers,
    domain=-5:5,
    samples=100,
    axis lines*=left,
    width=2.75cm, height=2.75cm,
    ymin=0,
    ymax=0.45,
    xtick=\empty,
    ytick=\empty,
    enlargelimits=false,
    clip=false,
    axis on top,
    grid = major
  ]
    
    % First Normal Distribution
    \addplot [
        very thick,
        orange!70!red
    ] {exp(-0.5*((x-1)/1.2)^2)/sqrt(2*pi*1.2^2)};
  \end{axis}

  % Targets 
  \plotElement{6.75}{-3.25}{Targets}{orange}{above}

  % Arrows 
  \draw[->, thick, draw=black] (-3,-3.25) -- node[below, pos=0.3] {$\mathcal{M}$} (e);
  \draw[->, thick, draw=black] (e) -- node[below, pos=0.7] {$\mathcal{H}_1$} (1.25,-3.25);
  \draw[->, thick, draw=black] (2.25,-3.25) -- node[below, pos=0.5] {$\hat{Z}$} (3.5,-3.25);
  \draw[<-, thick, draw=black] (5.0,-3.25) -- node[below, pos=0.5] {$Z$} (6.25,-3.25);
  \draw[thick, draw=black] (4.25,-4) -- (4.25,-4.6);
  \draw[thick, draw=black] (0.25,-4.6) -- node[above, pos=0.5] {$\mathcal{L}_1$} (4.25,-4.6);
  \draw[<-, thick, draw=black] (e) -- (0.25,-4.6);

  \node at (-4, -4.5) {Stage 1 Model};
  \draw[dashed] (-5.5, -4.75) -- (8, -4.75);
  \node at (-4, -5.05) {Stage 2 Model};
  
  %%% Stage 2 %%% 

  % Structural Model 
  \node at (-1, -7.1) {Structural Model};
  \draw[thick] (-1.75, -6.85) rectangle (-0.25, -5.35);

  % Second Plot
  \begin{axis}[
    name=plot2,
    at={(-1.55cm, -6.7cm)},
    anchor=south west,
    no markers,
    domain=-5:5,
    samples=100,
    axis lines*=left,
    width=2.75cm, height=2.75cm,
    ymin=0,
    ymax=0.5,
    xtick=\empty,
    ytick=\empty,
    enlargelimits=false,
    clip=false,
    axis on top,
    grid = major
  ]
    
    % Second Normal Distribution
    \addplot [
        very thick,
        cyan!70!blue
    ] {exp(-0.5*((x+1)/1.0)^2)/sqrt(2*pi*1.0^2)};

  \end{axis}

  % Estimates 
  \plotElement{1.75}{-6.1}{Estimates}{red}{below}

  % Error Metric 
  \node at (4.25, -7.1) {Error Metric};
  \draw[thick] (5, -6.85) rectangle (3.5, -5.35);
  \begin{axis}[
    name=plot3,
    at={(3.65cm, -6.7cm)},
    anchor=south west,
    no markers,
    domain=-5:5,
    samples=100,
    axis lines*=left,
    width=2.75cm, height=2.75cm,
    ymin=0,
    ymax=0.45,
    xtick=\empty,
    ytick=\empty,
    enlargelimits=false,
    clip=false,
    axis on top,
    grid = major
  ]
    
    % First Normal Distribution
    \addplot [
        very thick,
        orange!70!red
    ] {exp(-0.5*((x-1)/1.2)^2)/sqrt(2*pi*1.2^2)};
  \end{axis}

  % Targets 
  \plotElement{6.75}{-6.1}{Targets}{orange}{below}

  % Arrows 
  \draw[->, thick, draw=black] (e) -- node[left, pos=0.85] {$\mathcal{H}_2$} (-1,-5.35);
  \draw[->, thick, draw=black] (-0.25,-6.1) -- node[below, pos=0.5] {$\mathcal{F}$} (1.25,-6.1);
  \draw[->, thick, draw=black] (2.25,-6.1) -- node[below, pos=0.5] {$\hat{Y}$} (3.5,-6.1);
  \draw[<-, thick, draw=black] (5.0,-6.1) -- node[below, pos=0.5] {$Y$} (6.25,-6.1);
  \draw[thick, draw=black] (4.25,-5.35) -- (4.25,-4.95);
  \draw[thick, draw=black] (0.25,-4.95) -- node[below, pos=0.65] {$\mathcal{L}_2$} (4.25,-4.95);
  \draw[<-, thick, draw=black] (-1,-4.95) -- (0.25,-4.95);
  
\end{tikzpicture}
\caption{Proposed Model}
\label{fig:proposed-model}
\end{figure}

The proposed approach employs a two-stage estimation process, as illustrated in Figure \ref{fig:proposed-model}. In Stage 1, a language model $\mathcal{M}$ is trained to generate a DEV that {\it encodes} point predictions $\hat{Z}$ for the target outcomes $Z$ using the textual description of a product as input. To achieve this, a weight vector $\mathcal{H}_1$ is simultaneously optimized to project the DEV into the prediction space corresponding to these outcomes. For the empirical illustration of online car auction data, the target outcomes include metrics on bid values, the number of unique bidders, auction views, watchers, and whether the reserve price was met. Both $\mathcal{M}$ and $\mathcal{H}_1$ are trained with an error metric $\mathcal{L}_1$, ensuring that the DEVs effectively capture the necessary market information to serve as the foundation for subsequent econometric analysis. 

In Stage 2, a fully connected feedforward network, $\mathcal{H}_2$, {\it decodes} the market information embedded in the DEVs from Stage 1 into the parameter space of the structural model $\mathcal{F}$, which produces estimates $\hat{Y}$ for the target market outcomes $Y$, which may be a subset of the target market outcomes $Z$ from Stage 1. This second stage is crucial for enabling robust counterfactual analysis, as it explicitly constrains the model to use the assumed data generating process to predict the observed target outcomes. In the empirical illustration, these estimates are obtained using semi-nonparametric methods to model probability density functions over consumer valuations and market size for each auction vehicle's description. The network $\mathcal{H}_2$ is trained with an error metric $\mathcal{L}_2$, optimizing over the predicted outcomes from $\mathcal{F}$ while ensuring adherence to the specified structural relationships. 

This {\it encoder-decoder} approach compresses relevant market information into DEVs during Stage 1 and then transforms it into optimized estimates within the econometric framework in Stage 2. By fine-tuning the language model in Stage 1, the model learns subtle patterns in product descriptions that reflect consumer preferences and the value generated by these products, establishing a sophisticated mapping between unstructured textual features and market outcomes. The structural constraints imposed in Stage 2 then ensure that counterfactual analyses remain grounded in economic theory, making the predictions more reliable when simulating market interventions or policy changes. This approach overcomes key limitations of traditional hedonic models by eliminating the need for \emph{a priori} assumptions, enhancing flexibility and scalability across diverse markets and product categories, and improving the granularity of analysis to detect nuances in consumer preferences and market dynamics. 

To illustrate the proposed model, we collected transactional data from a large US online auction house for used vehicles. Using time-stamped and user-identifiable (yet anonymous) bids on approximately 75,000 car auctions from 2014-2023, we train both stages of this model and test its ability to generalize to previously unseen descriptions on a holdout sample of 5,000 auction observations, including an ablation study, or ``zero shot'' test, with an unusual make and model of vehicle -- the DMC DeLorean. 

To benchmark the model against familiar econometric baseline, we also estimate a conventional pooled-OLS hedonic specification and show the two-stage attention-based language model's superior prediction performance. Additionally, we validate the need for a two-stage model framework by training an alternate ``direct'' specification where we ignore the first stage $\mathcal{H}_1$ estimation and simply train $\mathcal{M}$ and $\mathcal{H}_2$ simultaneously. Although this direct estimation procedure predicts the training data as well as the proposed two-stage approach, we demonstrate it severely underperforms over the unseen validation set (i.e. worse predictions than taking a static average). % We conjecture the reason is that the economic value associated with the natural language description does not converege and instead keeps shifting as the structural model estimates are simultaneously calibrated. See Section \ref{sec:discussion}.

Our ability to offer these results is based on two recent developments. First, the scaling of machine learning algorithms has generated powerful emergent abilities to parse semantic meaning from natural language. In the context of natural language processing, an attention-based language model's architecture allows it to establish nonlinear relationships between various components of a text sequence \citep[][see \cite{qiu2020pre} for a survey]{mikolov2013linguistic}. After being trained on large swaths of the Internet to predict missing or ``masked'' text elements, these models can later be fine-tuned for alternative natural language applications \citep{devlin2018bert}. If tasked with predicting market outcomes such as prices from the natural language descriptions of products, as done here, training the language model will convert its initial semantic representations of goods to demand representations that are suitable for downstream econometric estimation. Second, the benefit of the abundance of naturally occurring market data, such as online auctions, that can offer strong signals of individual consumer valuations while satisfying the training requirements of a language model. 

The managerial implications of integrating machine learning with economic modeling extend beyond improved prediction; they redefine how firms generate causal insights for strategic decision making \citep{athey2018impact}. Language models unlock vast new data sources, such as product descriptions and written consumer feedback. However, their outputs must be structured within econometric models that enable reliable counterfactual analysis. The contribution of this paper is that the approach bridges the aforementioned gap by embedding language-model-driven demand signals into a structural model, and thus allowing managers to move beyond correlation-based forecasting to derive actionable, policy-relevant insights. This is particularly critical in digital marketplaces, where firms must anticipate demand shifts, optimize platform design, and refine dynamic pricing strategies in response to evolving market conditions \citep{mehta2021sell}. By integrating machine learning representations with economic inference techniques, the model ensures that machine-driven recommendations align with underlying economic forces, enhancing their applicability for real-world inferences. However, realizing this potential requires overcoming key challenges, including bias mitigation, interpretability, and model robustness across different market contexts \citep{van2022overcoming}. As firms increasingly rely on machine learning for decision support, the approach offers a framework for merging algorithmic flexibility with economic inference, establishing a foundation for machine-powered business strategy. Thus, our ambition is for this paper and its proposed model to serve the calls of \cite{abbasi2024pathways} and \cite{sarker2025advancing}.

The rest of the paper proceeds as follows. Following an overview of the existing literature in Section \ref{sec:lit_review}, we provide a more detailed description of the model framework in Section \ref{sec:model}, and the technical model training in Section \ref{sec:training}. In Section \ref{sec:data} we provide an overview and summary statistics of the online auction data, and then in Section \ref{sec:discussion} provide an empirical illustration of the proposed model framework, and discuss the numerical results and counter-factual analysis from the online auction data. Finally, in Section \ref{sec:conclusion}, we conclude with some remarks and ideas for future extensions.

Throughout this paper, we adhere to a notational convention in which uppercase letters indicate random variables, lowercase letters represent their realizations, Greek letters are used to denote model parameters, and cursive letters denote a model-specific function. As the ensuing empirical illustration is based on vehicles, we use the term ``product'' instead of ``good,'' but stress that the presented framework equally applies to ``services''. 

\section{Literature Review} \label{sec:lit_review}

With the creation of the transformer architecture \citep{vaswani2017attention}, attention-based language models have achieved unprecedented capabilities in representing the semantic meaning of natural language. Initially developed for natural language translations, these models have since demonstrated broad applicability across diverse domains, including text generation \citep{radford2019language}, protein structure prediction \citep{jumper2021highly}, medical image processing \citep{chen2021transunet}, writing advertising copy \citep{Chen2024}, privacy policy changes \citep{lin2024automated}, qualitative research \citep{zhou2024can}, and prediction on course and instructor evaluation \citep{wang2025predicting}. More recently, these advancements have led to a surge in the use of language models for empirical research in economics, business, and management, where textual data is increasingly leveraged to make economic insights on consumer preferences, firm strategies, and market and business dynamics \citep{gentzkow2019text, Berger2020, ash2023text, deKok2025}. 

Although language models have proven effective in areas such as sentiment prediction in customer service \citep{Puranam2021}, personality detection \citep{yang2023getting}, product reviews \citep{ma2023beyond}, aggregate consumer decisions \citep{qiu2023consumer}, and financial disclosures \citep{mccarthy2024fin}, their potential for demand estimation remains underexplored. Recent advances suggest that language models may provide a useful tool for extracting latent demand signals from natural language descriptions of goods and services, offering new possibilities for new approaches to demand estimation in a variety of market settings \citep{Timoshenko2019,Liu2023,wang2024large, chen2025conversation}. 
%\citep{Timoshenko2019,Wang2023,Liu2023,wang2024large, chen2025conversation}. 

The integration of machine learning into econometric methods has led to advances in predictive accuracy and modeling flexibility, especially in settings where traditional methods face limitations \citep{athey2019machine}. Recent research has leveraged deep learning techniques to capture complex relationships between product attributes and consumer preferences, incorporating high-dimensional representations derived from structured and unstructured data sources \citep{adam2024machine, aceves2024mobilizing}. Language models, in particular, have been used to explore individual preferences and risk tolerances in a variety of settings \citep{dillion2023can, wu2023large, goli2024frontiers, zhu2024language}.
%\citep{dillion2023can, wu2023large, goli2024frontiers, zhu2024language, Netzer2019}.

Although these methods improve predictive performance, a key challenge remains their lack of structural coherence and interpretability.\footnote{See \cite{olah2020zoom, elhage2022toy, templeton2024scaling} for language model interpretability methods.} Models that directly estimate economic primitives within a structural framework are typically better suited for counterfactual analysis \citep{verma2024counterfactual}, and improve performance on ``zero-shot'' (i.e. novel) problems \citep{wei2021finetuned}, yet many machine learning approaches lack this foundation. Recognizing this, recent work has explored hybrid approaches that integrate machine learning representations into structural models, using deep embeddings as inputs to traditional econometric frameworks \citep[e.g.][]{vafa2024estimating}. These approaches suggest a path forward for demand estimation methods that combine the flexibility of machine-driven representations with the interpretability and counterfactual validity of structural econometric models. 

\section{Econometric and Machine Learning Framework}\label{sec:model}

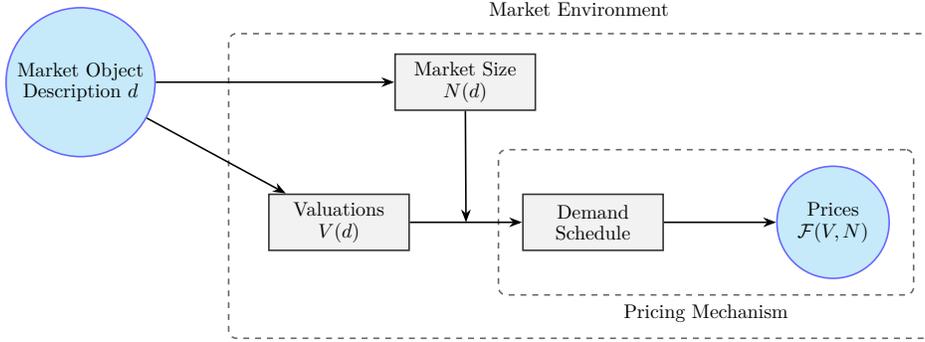
\begin{figure}
\begin{center}
\resizebox{0.75\textwidth}{!}{%
\begin{tikzpicture}[
  % Styles
  concept/.style={
    draw=blue!60, fill=cyan!20, thick,
    circle, minimum size=2cm, align=center
  },
  process/.style={
    draw=black!80, fill=gray!10, thick,
    rectangle, minimum width=2.5cm,
    minimum height=1cm, align=center
  },
  arrow/.style={-Stealth, thick},
  group_box/.style={
    draw=black!60, dashed, thick,
    rounded corners, fill=white, fill opacity=0.7
  }
]

  % Primary nodes
  \node[concept]                           (D)  {Market Object\\Description $d$};
  \node[process,right=2cm of D, yshift=-2.5cm] (V)  {Valuations\\$V(d)$};
  \node[process,right=4.25cm of D]         (N)  {Market Size\\$N(d)$};
  \node[process,right=2cm of V]            (DS) {Demand\\Schedule};
  \node[concept,right=2cm of DS]           (P)  {Prices\\$\mathcal{F}(V,N)$};

  % Arrows and intersection point
  \draw[arrow] (D) -- (V);
  \draw[arrow] (D) -- (N);
  \draw[arrow] (V) -- coordinate[midway] (VDSmid) (DS);
  \draw[arrow] (DS) -- (P);
  \draw[arrow] (N) -- (VDSmid);

  % Background grouping boxes
  \begin{scope}[on background layer]
    % Outer "Market Environment" box
    \node[group_box, fit=(V)(N)(P),
          inner xsep=20pt, inner ysep=20pt, yshift=-10pt,
          label={[yshift=5pt]above:Market Environment}] {};
    % Inner "Pricing Mechanism" box (excludes Demand Schedule)
    \node[group_box, fit=(DS)(P),
          inner xsep=12pt, inner ysep=8pt,
          label={[yshift=-1pt]south:Pricing Mechanism}] {};
  \end{scope}

\end{tikzpicture}
}
\end{center}
\caption{Assumed Data Generating Process}
\vspace{-0.5em}
\begin{footnotesize}
    \textit{Note:} Market observables are in circles, latent variables are in rectangles. 
\end{footnotesize}
\label{fig:data-flow-1}
\end{figure}

The framework for the proposed model begins with a set of assumptions over the data generating process on market demand, outlined in Figure \ref{fig:data-flow-1}. Within a given market environment, the introduction of a product with description $d$ generates, to an exposed population, two stochastic demand primitives that together produce an aggregate demand schedule. The first primitive of interest is $V(d)$, a continuous random variable that produces {\it i.i.d.} realizations on the {\it willingness to pay} that consumers have for $d$. The corresponding distribution function is given by $F_{V}(v)=\text{Pr}\{V(d)\le v\}$, with support on $\mathbb{R}$. The second primitive of interest is $N(d)$, a discrete random variable that represents the {\it number of potential consumers} a description $d$ attracts (i.e., market size). Likewise, this is defined by the distribution function $F_{N}(n)=\text{Pr}\{N(d)\le n\}$, with support on $\mathbb{N}$. To generate a demand schedule, we assume that the realization in market size $n$ determines the set of realizations over valuations $\{v_1,v_2,\dots,v_n\}$. Without loss of generality, we assume that the valuations are ordered, i.e. $v_n\leq \cdots \leq v_2 \leq v_1$. For simplicity, we assume that all consumers have unitary demand, and that $V(d)$ and $N(d)$ are independent for all $d$, i.e. $f_{V,N}(v,n)=f_{V}(v)f_{N}(n)$. 

Once valuations are realized, consumers engage with the market's pricing mechanism, which can be a function of the market structure and available information, to produce observable prices. In the context of the empirical application of online vehicle auctions, the structural model $\mathcal{F}$ follows the English auction format, which implies that demand for a vehicle is stochastically determined when it is up for auction. We do not predetermine the dynamics of $V(d)$ and $N(d)$ or the conditions for their realization. Instead, we leverage the language model to uncover these dynamics based on the vehicle's description, imposed structural model, and available market data. Next, we discuss the two-stage estimation method for the distributions of $V$ and $N$, as presented in Figure \ref{fig:proposed-model}.

\subsection{Encoding Economic Information in Market Object Descriptions} 

The proposed estimation approach uses an encoder-only language model to estimate the demand characteristics of a market object from the features represented in its description $d$. For this task, $d$ is broken down into a vector of the language model's most basic text units, called \emph{tokens}, which we denote as $\token$. Tokens are integer indices of words, subwords (words broken into smaller parts), or characters. To convert text to tokens, the language model is accompanied by a designated \emph{tokenizer}, which is a function $\mathcal{D}$ that converts a text string to a vector of integer values within the language model's vocabulary set, $\mathcal{D}: d \rightarrow \vec{\token}$. We denote the length of the token vector by $\tokentotal$. A pedagogical example of tokenization is provided in Table \ref{tab:tokens-1} using the tokenizer we employ in the empirical section of this study, which has a vocabulary set of approximately 30,000 distinct tokens \citep{li2023gte}. 

\begin{table}[t]
\centering 
\caption{Pedagogical Example of Tokenization and Demand Embedding Vectors} 
\resizebox{0.85\textwidth}{!}{
\begin{tabular}{l*{12}{c}}
\hline
% \textbf{Description} $d$: & 
% \tokenizertkn{\colorone}{This} & \tokenizertkn{\colortwo}{ 1974} & \tokenizertkn{\colorthree}{ BMW} & \tokenizertkn{\colorfour}{ 3} & \tokenizertkn{\colorzero}{.} & \tokenizertkn{\colorone}{0} & \tokenizertkn{\colortwo}{CS} & 
% \tokenizertkn{\colorthree}{ was} & \tokenizertkn{\colorfour}{ refurb} & \tokenizertkn{\colorzero}{ished} \\ 
\textbf{Description} $d$ & \texttt{[CLS]} & This & 1974 & BMW & 3 & . & 0 & CS & was & refurb & ished & ... \\ 
\textbf{Tokens} $\mathcal{D}: d \rightarrow \vec{\token}$ & 101 & 713 & 15524 & 8588 & 155 & 4 & 288 & 6842 & 21 & 17880 & 6555 & ... \\
\textbf{DEVs} $\mathcal{M}: \vec{\token}\rightarrow \{\vec{e}_j\}$ & $\vec{e}_0$ & $\vec{e}_1$ & $\vec{e}_2$ & $\vec{e}_3$ & $\vec{e}_4$ & $\vec{e}_5$ & $\vec{e}_6$ & $\vec{e}_7$ & $\vec{e}_8$ & $\vec{e}_9$ & $\vec{e}_{10}$ & ... \\
\hline
\end{tabular}
}
% \vspace{-0.5em}
\label{tab:tokens-1}
\end{table}

Attention-based language models process tokens through multiple layers of self-attention mechanisms, allowing each token to encode contextual information from the entire text sequence. This mechanism enables the model to evaluate the relevance of each token by considering its relationship with all other tokens \citep[see][]{turian2010word,mikolov2013linguistic,vylomova2015take}. To illustrate, in the ensuing online vehicle auction application, the model specifically encodes relevant market demand information associated with each object, e.g., the encoding for the tokens for ``manual transmission'' depends on the context provided by surrounding tokens, such as whether tokens representing ``Honda'' or ``Lamborghini'' are present.

Mathematically, each token $\token[j]$ in a textual description $d$ is individually mapped to a $q$-dimensional real-valued vector $\vec{e_j} \in \mathbb{R}^{q}$, labeled \emph{Demand Embedding Vectors (DEVs)}, $j=0,\dots,\tokentotal$,
\begin{equation*}
\mathcal{M}: \vec{\token}\rightarrow \{\vec{e}_j\}_{j=0}^{\tokentotal}
\end{equation*}
where the initial token $\token[0]$ is a special classification token \texttt{[CLS]} placed at the beginning of every token sequence. After processing through the self-attention layers, the \texttt{[CLS]} token aggregates information across all tokens, acting as an information-dense fingerprint that captures the overall context of the input description \citep{li2020sentence}. Thus, the language model $\mathcal{M}$ produces both token-specific DEVs and a comprehensive DEV via the \texttt{[CLS]} token, summarizing the collective market-demand context embedded in the description, see Table~\ref{tab:tokens-1}. 

\subsection{Stage 1 Estimation: Description-Level Representations of Market Outcomes}

The goal of the first stage estimation is to, for a given product description $d$, train $\mathcal{M}$ to \emph{encode} the relevant market information of the product into its \texttt{[CLS]} token's DEV $\vec{e}_0$. To do this, we append a fully connected network head $\mathcal{H}_1$ to $\mathcal{M}$ so that it projects this DEV to $m$ real-valued target outcomes of interest, $\hat{y}_{j} \in \mathbb{R}^{m}$. For example, in the case of vehicle auctions, the outcomes could include the submitted bid values, the number of active bidders, and how many times the auction was viewed. This approach leads to the following proposition, which reformulates a well-established result from the computer science literature in the context of the economic application. 

\medskip 
\noindent\textbf{Proposition 1.} Let $D=\{(d,\vec{y})\}$ be a set of text descriptions $d$ and associated economic information vectors $\vec{y}\in\mathbb{R}^m$. For any error tolerance $\epsilon > 0$, there exists a language model $\mathcal{M}:D\rightarrow\mathbb{R}^{q}$, and a decoding function $\mathcal{H}_1:\mathbb{R}^{q}\rightarrow \mathbb{R}^m$, such that 
\begin{equation*}
    \sup_{(d,y)\in D} \|y - \mathcal{H}_1\circ\mathcal{M}\|_2 < \epsilon. 
\end{equation*}
\noindent\textbf{Proof.} \emph{See Appendix \ref{app:proofs}}
\medskip 

Drawing on established universal approximation theorems in neural networks, this result simply restates that any information purposefully embedded in the language model's representation space is recoverable with arbitrary precision through an appropriate decoding function. Therefore, we apply this approach to generating predictions $\hat{y}_{i}$ on target outcome $i$ with the \texttt{[CLS]} token's DEV, producing a set of $m$ observed market outcomes $\{y_1(d), y_2(d), \dots, y_m(d)\}$ with the estimated decoder, 
\begin{equation}\label{eq:h-1}
    \hat{\mathcal{H}}_1: \vec{e}_0 \rightarrow \{\hat{y}_{i}\} \quad \text{for all } i \in \{1, \dots, m\}. 
\end{equation}
That is, the purpose of the $\hat{\mathcal{H}}_1$ head is to \emph{decode} the high-dimensional DEV representation of the \texttt{[CLS]} token into predictions on the target outcomes. 

We optimize the estimated parameters in $\hat{\mathcal{M}}$ and $\hat{\mathcal{H}}_1$ with a loss function that minimizes the mean squared error (MSE) between the estimated market outcomes $\{\hat{y}_1(d), \hat{y}_2(d), \dots, \hat{y}_m(d)\}$ and the true observed market outcomes $\{y_1(d), y_2(d), \dots, y_m(d)\}$,
\begin{equation}\label{eq:stage1-loss} 
    \mathcal{L}_1=\frac{1}{|D|}\sum_{d\in D}\sum_{i=1}^{m}\left[y_i(d)-\hat{y}_i(d)\right]^2
\end{equation}
where $D$ represents the set of training samples. The loss function $\mathcal{L}_1$ ensures that the model learns to produce accurate point predictions by reducing the squared deviations between the predicted and actual market outcomes. We label the ensuing models, with the optimized parameters minimizing $\mathcal{L}_1$, by $\hat{\mathcal{M}}^{*}$ and $\hat{\mathcal{H}}_{1}^{*}$.

While Stage 1 produces a representation of each description through the information-dense embedding of the \texttt{[CLS]} token, it does not by itself yield direct inference on the structural parameters governing market outcomes. To address this, the second-stage estimation process projects the \texttt{[CLS]} embedding—summarizing the full input description—into the parameter space of an econometric model. This enables a more interpretable and structured analysis of demand primitives. 

\subsection{Stage 2 Estimation: Description-Level Representations to Demand Primitives}

The objective of Stage 2 is to estimate two latent demand primitives of our structural model: the valuation distribution $F_V(v)$ and the market size distribution $F_N(n)$. While the machine learning approach in Stage 1 successfully encodes market outcomes into DEVs derived from product descriptions' \texttt{[CLS]} tokens, these predictions alone do not identify the structural economic forces driving demand; for further discussion, see \cite{mullainathan2017machine, iskhakov2020machine}. By estimating the respective demand primitives via a structural model, we constrain the prediction process to follow the assumed data generating process, which is essential for performing robust counterfactual analyses. This structural approach enables us not only to infer the underlying relationships generating consumer behavior but also to make reliable predictions for unseen applications, such as new product descriptions or market structures, by ensuring that all counterfactual scenarios remain consistent with the economic theory embedded in our model.

To enable a general and flexible representation of the valuation distribution, we employ a semi-nonparametric approach developed by \cite{gallant1987semi}, \cite{fenton1996convergence}, and \cite{fenton1996qualitative}. Specifically, the density function $f_{V}(v)$ is approximated by,
\begin{equation} \label{eq:hermite-1} 
\hat{f}_{V}(v) = c\left[1+\sum_{k=1}^{\kappa}\alpha_{k}\left(\frac{v-\mu}{\sigma}\right)^{k}\right]^2\varphi(v\mid \mu,\sigma)
\end{equation} 
where $\varphi(v\mid \mu,\sigma)$ is the Gaussian density component that imparts location ($\mu$) and scale ($\sigma$) properties, and, for a given level $\kappa\in \mathbb{N}$, a Hermite polynomial expansion capturing the deviations from normality through the weighted terms $\alpha_{k}$. The squaring of the expansion guarantees non-negativity, and the normalizing constant,
\begin{equation*}
c = \left[\int_{-\infty}^{\infty} \left(1+\sum_{k=1}^{\kappa} \alpha_{k} \left(\frac{z-\mu}{\sigma}\right)^{k}\right)^2 \varphi(z\mid \mu,\sigma)dz \right]^{-1}
\end{equation*} 
ensures that $\hat{f}_V(v)$ integrates to one, and thus a well-defined probability density function. This specification yields a parameter space for the valuation distribution of $\vec{\lambda}_V=\{\alpha_1,\dots,\alpha_\kappa,\mu,\sigma\}$. 

For the market size distribution, we adopt a semi-nonparametric approach that uses the softmax function to construct a probability mass function. Specifically, the parameter space for the market size distribution $\vec{\lambda}_N=\{\rho_{\underline{n}},\dots,\rho_{\overline{n}}\}$, consists of likelihood values $\rho_r$ on the discretized grid of potential bidder counts $B=\{\underline{n},\dots,\overline{n}\}$, $B\subseteq \mathbb{N}$. The likelihood values are then transformed via the softmax function, which exponentiates and normalizes them so that they are nonnegative and sum to one. This transformation yields a probability mass function for each $n$ in the defined grid. That is, the probability mass function $f_{N}(n)$ is approximated by, for $n\in\{\underline{n},\dots,\overline{n}\}$,
\begin{equation} \label{eq:softmax-1} 
    \hat{f}_N(n) = \frac{\exp(\rho_n)}{\sum_{r\in B}\exp(\rho_r)}
\end{equation}
This formulation not only captures the inherently discrete nature of market participation but also allows for a flexible, data-driven estimation of the underlying bidder distribution. 

Following \cite{farrell2020deep} on deep network architecture, our Stage 2 estimation approach integrates a neural network with semi-nonparametric density estimation techniques. Specifically, we append a fully connected network head, $\mathcal{H}_2$, to the fine-tuned language model $\mathcal{M}^*$ from Stage 1. For each product description $d$, $\mathcal{M}^*$ produced the \texttt{[CLS]} token's DEV, $\vec{e}_0$, which is then \emph{decoded} by $\mathcal{H}_2$. This network head maps the classification DEV into the parameter spaces of both distributions, 
\begin{equation*} 
\mathcal{H}_2: \vec{e}_0 \rightarrow \lbrack\vec{\lambda}_V, \vec{\lambda}_N\rbrack 
\end{equation*}

The dual-head architecture ensures that although both distributions are derived from the same DEV, each branch is tailored to its respective estimation task. By mapping the DEVs into these conditional density spaces, $\mathcal{H}_2$ effectively decodes the market-relevant signals encoded in Stage 1 into coherent, interpretable demand estimates well-defined for structural analysis. 

Finally, we include a note on how one should seek to train $\mathcal{H}_2$ (i.e., estimate the model parameters). It is necessary to specify how the latent demand primitives generate the observed target outcomes through the market-specific pricing mechanism, as depicted in Figure \ref{fig:data-flow-1}. That is, the calibration of $\mathcal{H}_2$ is context-dependent on the data generating process. Motivated by the ensuing empirical analyses, we illustrate by considering the English auction format.

\subsubsection{The English Auction Pricing Mechanism and Equilibrium Bidding.}

In an English auction, participating bidders sequentially submit incremental bids until no bidder is willing to exceed the current highest bid and the auction concludes. It should be fairly intuitive that, a rational bidder $j$, with private valuation $v_j$, should continue to incrementally bid $b_j^+$, as long as their surplus $v_j - b_j^+$ is non-negative. Hence, bidders' weakly dominant strategy is to bid up to their valuation, $b_j^+\leq v_j$, and, conditional on the seller's reserve price $p_R$ being met, the final transaction price is determined by the \emph{second highest valuation} plus the bid increment submitted by the winning bidder. To simplify notation, denote bidder $j$'s \emph{last bid} as $b_j$, $j=1,\dots, n$. Given ordered valuations, it follows that $b_j=v_j$, for $j=2, \dots, n$, while $b_1=v_2+b_{inc}$, where $b_{inc}$ is the minimum bid increment. That is, for a subset of bidders, it is assumed the last (maximum) bid submitted represents their valuation. Note, in the English auction the winning bidder's valuation is censored and only known to exceed the second highest valuation with the minimum bid increment. We apply this revelation of valuations through the equilibrium bid strategy to the auction data. 

While the equilibrium bidding strategy for the English auction is robust to several complicating factors, including risk tolerance, the number of bidders \citep{wolfstetter1996auctions}, and other bidders strategies \citep{lu2019information}, there are a couple of potential shortcomings. First, it is unable to point identify the necessary order statistics in the presence of jump bidding, i.e. bidding much more than the minimum bid increment; see \cite{avery1998strategic,easley2004jump}. If there are large bid gaps due to jump bidding then assuming $b_j=v_j$ will produce biased estimates. Second, this is also a concern if participation is temporally intermittent, or sparse, or subject to rapid counter-bidding. For instance, if two bidders aggressively counter-bid each other, then the price trajectory might escalate quickly such that other bidders' last bids are far below their respective valuations. Finally, if there are multiple ongoing, or sequential, auctions of identical or similar objects, then bidders may strategically not bid up to their valuation in a given auction, in anticipation of maximizing their expected surplus over a set of auctions. These shortcomings may thus introduce estimation bias on the order statistics of the $2^\text{nd}, 3^\text{rd}, 4^\text{th}, \dots$ highest valuations. We address these concerns in our inclusion criteria for the auction data discussed in the empirical section of the paper. 

\subsubsection{Defining the Structural Model $\mathcal{F}$ for an English Auction.}

Given that the equilibrium bid strategy can serve as reasonable proxies for at least some bidders' private valuation, we use the approximated density and probability mass functions, $\hat{f}_V(v)$ and $\hat{f}_N(n)$, to predict the order statistics of the observed bids in the auctions. Specifically, recall that the number of bidders $n$ is drawn from $F_{N}(n)$, and bidders' valuation are drawn i.i.d. from $F_{V}(v)$. This implies that conditional on a realized market size $n$, the estimated density function for the order statistics of the top $j \leq n$ valuations, is given by, for $j^\prime=j,\dots,2,1$, 
\begin{equation}\label{eq:joint-density-1}
    \hat{f}_{V_{(j^\prime)}}(v\vert n) = \frac{n!}{(j^\prime-1)!(n-j^\prime)!} \hat{f}_{V}(v)(\hat{F}_{V}(v))^{n-j^\prime}(1-\hat{F}_{V}(v))^{j^\prime-1}.
\end{equation}

Using Equation \eqref{eq:joint-density-1}, the structural model $\mathcal{F}$ numerically computes, conditional on market size $n$, the expectation of the top $j$ order statistic by,  
\begin{equation*} % \label{eq:conditional-expectation}
\hat{E}\left[V_{(j^\prime)} \mid n\right] = \int_{-\infty}^{\infty} v \hat{f}_{V_{ (j^\prime)}}(v\vert n) \, dv.
\end{equation*}
To compute the unconditional expectation, we average over the distribution of $n$, for $j^\prime=j,\dots,2,1$, 
\begin{equation*} % \label{eq:unconditional-expectation}
\hat{E}_d\left[V_{(j^\prime)}\right] = \sum_{n = \underline{n}}^{\overline{n}} \hat{f}_{N}(n) \, \hat{E}\left[V_{(j^\prime)} \mid n\right].
\end{equation*}
Finally, we collect these expectations into a vector representing the predictions of the $j$ largest bids, 
\begin{equation}\label{eq:expected-vector}
\left[\hat{b}_1, \hat{b}_2,\dots, \hat{b}_j\right] = \left[\hat{E}_d\left[V_{(1)}\right], \, \hat{E}_d\left[V_{(2)}\right], \, \dots, \, \hat{E}_d\left[V_{(j)}\right]\right].
\end{equation}

To train $\mathcal{H}_2$, we optimize its parameters to ensure that the estimated valuation and market size distributions generate expected bid predictions via $\mathcal{F}$ that align with the observed bid results. As previously discussed, the largest bid $b_1$ is biased downward due to the bidding strategies induced by the English auction. Therefore, $\mathcal{H}_2$ is trained to estimate valuations $\{V_2, \dots, V_j\}$, using the bids $\{b_2, \dots, b_j\}$, according to their computed order statistics from $\mathcal{F}$. We do this by minimizing the mean squared error (MSE) between the predicted bids $\hat{b}$ and the observed bids $b$ for the bids that can proxy for bidder valuations. The loss function is given by,
\begin{equation} \label{eq:stage2-loss}
    \mathcal{L}_2 = \frac{1}{|D|} \sum_{d \in D} \sum_{j^\prime=2}^{j} \left[ b_{j^\prime}(d) - \hat{b}_{j^\prime}(d) \right]^2
\end{equation}
where $D$ represents the set of training auctions. This objective ensures that the estimated distributions $\hat{F}_V(v)$ and $\hat{F}_N(n)$ generate bid predictions that match empirical observations. Similar to above in Stage 1, we denote $\mathcal{H}_2^*$ as the model with the optimized parameters that minimize $\mathcal{L}_2$.

Although the estimation framework for $\mathcal{F}$ is tailored for English auctions, our broader approach can accommodate alternative market settings. For example, in posted-price markets, firms set prices based on expected residual demand rather than competitive bidding. Similarly, in bargaining environments, price formation is influenced by strategic negotiation between buyers and sellers. To extend our methodology to such settings, the structure of $\mathcal{F}$ must be adjusted to reflect the equilibrium conditions that govern the formation of prices in these alternative markets. 

\section{Language Model Training and Structural Model Estimation}\label{sec:training}

The foundation of our training framework is the base mGTE encoder-only language model introduced by \cite{li2023gte}. mGTE is an advanced variant of the {\it Bidirectional Encoder Representations from Transformers} (BERT) model that demonstrated how a transformer-based language model could encode information from unstructured text \citep{devlin2018bert}. In our application to online auction data, we use the base mGTE model to initialize our language model $\mathcal{M}$, which facilitates a two-stage training process to extract structured demand representations and estimate the economic primitives assumed to be governing the observed auction outcomes. Table \ref{tab:training-method-1} summarizes training setup and the hyperparameter settings used in both stages. 

For the empirical analysis, only auctions that met the time-based inclusion criteria, which required that the last (largest) bids from five unique bidders be submitted within 3 hours of the auction's end time, were used in the training and validation sets. This resulted in a total of 74,605 auctions for training, 5,000 auctions withheld (i.e. not seen by the model during training), and 93 ``zero-shot'' auctions for the DMC DeLorean for validation analysis. More details on the data are provided below in Section \ref{sec:data}. 

\begin{table}[t]
\centering 
\caption{Model Training Setup and Hyperparameters}
\label{tab:training-method-1}
\resizebox{0.9\textwidth}{!}{
\begin{tabular}{lcc}
\hline\hline 
 & \textbf{Stage 1} & \textbf{Stage 2} \\ 
\textbf{Training Objective} & Encoding Demand & Estimating Structural Parameters \\
\hline \\ [-2.25ex] 
\textbf{Parameters} & $\mathcal{M}$ (mGTE) and $\mathcal{H}_1$ & $\mathcal{H}_2$ \\ 
\hline \\ [-2.25ex] 
\textbf{Targets} $\vec{y}$ & Bids $b_2, b_3, b_4, b_5$ & Bids $b_2, b_3, b_4, b_5$ \\ 
 & Auction Views &  \\ 
 & Auction Watchers &  \\ 
 & No. Active Bidders &  \\ 
 & Reserve Met Indicator $\mathbbm{1}[b_1 \geq p_R]$ &  \\ 
\hline \\ [-2.25ex] 
\textbf{Training Dataset} & \multicolumn{2}{c}{74,605 Auctions} \\ % Authentic and 298,792 Synthetic Descriptions
\textbf{Validation Datasets} & \multicolumn{2}{c}{5,000 Auctions and 93 ``Zero-shot'' Auctions} \\ 
\hline \\ [-2.25ex] 
\textbf{Optimizer} & AdamW & AdamW \\ 
Learning Rate & $3 \times 10^{-5}$ (max) & $3 \times 10^{-5}$ (max) \\ 
$\beta_1$ & 0.9 & 0.9 \\ 
$\beta_2$ & 0.999 & 0.999 \\ 
$\epsilon$ & $1 \times 10^{-8}$ & $1 \times 10^{-8}$ \\ 
\hline \\ [-2.25ex] 
\textbf{Scheduler} & Linear Decay w/ Warmup & Linear Decay w/ Warmup \\ 
Warmup Steps & 300 & 600 \\ 
\hline \\ [-2.25ex] 
\textbf{Training Iterations} & 5 epochs, batch size 64 & 5 epochs, batch size 32 \\ 
\hline 
\multicolumn{3}{l}{Note: Target values in $\vec{y}$ were perturbed with a small amount of Gaussian noise during} \\
\multicolumn{3}{l}{training to prevent over-fitting (aside from the reserve met indicator variable).}
\end{tabular}
}
\end{table}

\subsection{Stage 1: Encoding Demand Embedding Vectors}

The objective of Stage 1 is to train $\mathcal{M}$ to encode demand-relevant outcomes into the \texttt{[CLS]} embeddings generated from the auction items' descriptions. To project this token's demand embedding vector $\vec{e}_0$ from $\mathcal{M}$ into the auction outcome space, we append a fully connected neural network layer, $\mathcal{H}_1$. Specifically, $\mathcal{H}_1$ produces estimates for the log value of the second through fifth highest bid values ($b_2, b_3, b_4, b_5$), as well as the auction-level engagement metrics total views, watchers, and unique bidders. During training, a small amount of Gaussian noise was added to the true target values of these variables to prevent over-fitting of the model. Additionally, $\mathcal{H}_1$ predicts whether the auction's reserve price $p_R$ was met, denoted by the binary indicator $\mathbbm{1}[b_1 \geq p_R]$. Given the binary nature of this target outcome, we apply a sigmoid transformation to the prediction made by $\mathcal{H}_1$ to ensure its values are in the unit interval, $[0,1]$. Stage 1 is trained for 5 epochs using the AdamW optimizer with a maximum learning rate of $3 \times 10^{-5}$ and a linear warmup decay schedule. We employ a batch size of 64 and monitor the mean squared error (MSE) loss in equation \eqref{eq:stage1-loss} across all predicted outcomes in the training dataset. 

\subsection{Stage 2: Estimating Structural Demand Parameters}

The second stage of training refines the structured representations of demand learned in Stage 1 by estimating the latent distributions of consumer valuations and market size in our structural model $\mathcal{F}$. While Stage 1 provides direct predictions of auction outcomes, Stage 2 introduces a structural estimation framework that parameterizes the probability density functions of the demand primitives as defined in equations \eqref{eq:hermite-1} and \eqref{eq:softmax-1}, allowing for a deeper economic interpretation of how textual descriptions influence bidder behavior. 

In Stage 2, the demand embedding vector $\vec{e}_0$ from $\mathcal{M}^*$ is frozen and passed through a nonlinear transformation layer in $\mathcal{H}_2$. This transformation extracts a lower-dimensional representation optimized for structural inference. The transformation consists of two fully connected layers with hidden dimensions of 32 and 16, each followed by LayerNorm regularization and a SiLU activation function, which introduces smooth non-linearity to enhance feature extraction while maintaining numerical stability. Separate output heads within $\mathcal{H}_2$ then estimate the parameters of the valuation distribution $F_V(v)$ and market size distribution $F_N(n)$. The optimization objective stated in equation \eqref{eq:stage2-loss} for Stage 2 minimizes the discrepancy between the $2^\text{nd}-5^\text{th}$-largest bids submitted by unique bidders and those generated from the estimated joint order statistic distributions in equation \eqref{eq:joint-density-1}. As in the Stage 1 estimation, a small amount of Gaussian noise was added to these variables during training. 

Similar to \cite{liu2020preserving}, we use Sobol sequences, which are low-discrepancy quasi-Monte Carlo (QMC) sequences, to efficiently approximate the expectations of these order statistics. This approach provides more uniform coverage of the probability space compared to standard Monte Carlo sampling \citep{morokoff1995quasi,jackel2002monte}, and also allows for a more stable and sample-efficient numerical integration when computing bid expectations over the estimated valuation distribution. Stage 2 is trained using the AdamW optimizer with a learning rate of $3 \times 10^{-5}$ and a batch size of 32. The training process spans 5 epochs and incorporates a linear warmup schedule with 1,000 warmup steps.

Once trained, the full model consists of $\mathcal{M}^*$ and $\mathcal{H}_2^*$, which can be applied to any vehicle description. In the empirical section that follows, we validate the predictive accuracy of the model and perform counterfactual simulations to examine how different vehicle descriptions influence estimated demand.

\section{Online Vehicle Auction Data -- BringATrailer.com} \label{sec:data}

For the empirical application of our estimation method, we collected data from \href{bringatrailer.com}{bringatrailer.com} (BaT), one of the largest online vehicle auction marketplaces in the USA. %classic and exotic vehicles. 
BaT utilizes an ascending English auction mechanism with a secret reserve price $p_R$ for their vehicle auctions. The description of each auction vehicle is written by BaT staff following a stylized template that reviews the history of the vehicle, its current condition, any modifications or restorations, and its provenance. 

Each auction begins at \$1 (all prices in USD), typically with a seven day clock that eventually ends the auction. To minimize {\it snipe bidding}, BaT automatically extends the auction's clock by two minutes whenever a new bid is submitted within this threshold. When the auction concludes, the highest bid $b_1$ is privately compared to the vehicle's secret reserve price $p_R$, and automatically announced as to whether the reserve was met. If the seller's reserve price is met ($b_1\geq p_R$), then the winning bidder pays the seller the amount $b_1$. In addition, the winning bidder is charged a BaT transaction fee of 5\% of $b_1$ or \$250, whichever was greater, but no more than \$5,000, i.e. $\min(\max(.05\times b_1, \$250), \$5,000)$. If the seller's reserve price is not met ($b_1<p_R$), then BaT connects the highest bidder with the seller to attempt a privately negotiated a sale price. However, regardless of the outcome, the seller is charged an unrecoverable listing fee of \$99. 

Important to this study, all text-based communications and bids submitted during a given auction were publicly recorded in a chat section at the bottom of the auction's webpage. While presumably designed to enhance the social aspects of the auction site, and reduce some information asymmetries, our estimation methods benefit from having every submitted bid timestamped and user-specified (yet anonymous) by BaT. We sourced all market data from the information in these chat sections. 

\subsection{Descriptive Summary Statistics}

From July 2014 through December 2023, BaT held a total of 89,080 auctions, however only 79,698 of them met our inclusion criteria that the final bids from five unique bidders be submitted within 3 hours of the auction's end time. This decreased the likelihood that some high value bidders were not present toward the end of the auction. Additionally, recall that to mitigate snipe biding, BaT extended the end time by two minutes whenever a bid was submitted within the last two minutes. We summarize the characteristics of the auctions meeting the inclusion criteria in Table \ref{tab:summary-1}. 

Approximately 31.6\% of the auctions were managed by dealers, and the reserve prices were met in 77.2\% of cases. Bidder participation averaged 12.007 active bidders per auction, with a minimum of 1 and maximum 39 unique bidders. For bid values, the average winning bid across the auctions was \$44,452, with the highest recorded bid reaching \$5,360,000 and the lowest at \$105. When normalized for analysis with the highest bid set to 100\%, the second and third highest bids averaging 96.9\% and 94.1\%, respectively, of the winning bid. Notably, the third highest bids submitted by non-winning participants (i.e. those who did not ultimately submit the single largest bid) averaged slightly lower, at 87.8\% of the highest bid.

\begin{table}[t]
\centering 
\caption{Descriptive Statistics of sample BringATrailer.com data (July 2014 - December 2023)} 
\resizebox{0.75\textwidth}{!}{
\begin{tabular}{lcccc}
    \hline\hline 
    \textbf{79,698 Auctions} & \textbf{Mean} & \textbf{St. Dev.} & \textbf{Min.} & \textbf{Max.} \\
    \hline \\ [-2.25ex]
    Bidding Duration (Days)$^\dagger$ & 6.902 & .920 & .0097 & 21.054 \\ 
    $\mathds{1}[\text{Sold by Dealer}]$ & .316 & .465 & 0 & 1 \\ 
    $\mathds{1}[\max(b)\geq r]$ (Reserve Met) & .772 & .419 & 0 & 1 \\ 
    No. of Active Bidders & 12.007 & 4.522 & 1 & 39 \\ 
    Auction Views (1000s) & 14.209 & 9.782 & .698 & 358.191 \\ 
    Auction Watchers (1000s) & .726 & .349 & .060 & 6.105 \\ 
    \hline \\ [-2.25ex]
    \multicolumn{1}{l}{\textbf{Nominal Bid Values}} &  &  &  &  \\\cline{1-1}
     \\ [-2.25ex]
     Winning Bid (in 1000s) & \$44.452 & \$79.284 & \$.105 & \$5,360 \\ [0.75ex] 
    \multicolumn{1}{l}{\textbf{Normalized Bid Values}} &  &  &  &  \\\cline{1-1}
     \\ [-2.25ex] 
    1$^\text{st}$ Largest (Winning) Bid & 1 & n/a & 1 & 1 \\ 
    2$^\text{nd}$ Largest Bid & .969 & .056 & .000 & .999 \\  
    3$^\text{rd}$ Largest Bid From... &  &  &  &  \\ 
    \hspace{0.5cm}...All Bidders & .941 & .083 & .000 & .998 \\ 
    \hspace{0.5cm}...Non-winning Bidders & .878 & .120 & .000 & .998 \\ 
    All Bids$^\dagger$ & .662 & .261 &  &  \\ 
    All Unique Bidders' Max. Bids$^\dagger$ & .615 & .295 &  &  \\ 
    \hline \\ [-2.25ex]
    \textbf{Fraction of Bids} &  &  &  &  \\ 
    \textbf{Submitted in Final...} & \textbf{5 Min.} & \textbf{15 Min.} & \textbf{1 Hr.} & \textbf{24 Hr.} \\
    \hline \\ [-2.25ex] 
    1$^\text{st}$ Largest (Winning) Bid & 1 & 1 & 1 & 1 \\ 
    2$^\text{nd}$ Largest Bid & .810 & .854 & .901 & .966 \\ 
    3$^\text{rd}$ Largest Bid From... &  &  &  &  \\ 
    \hspace{0.5cm}...All Bidders & .738 & .794 & .849 & .935 \\ 
    \hspace{0.5cm}...Non-winning Bidders & .835 & .860 & .891 & .948 \\ 
    All Bids$^\dagger$ & .160 & .342 & .470 & .633 \\ 
    All Unique Bidders' Max. Bids$^\dagger$ & .190 & .270 & .358 & .517 \\  [0.25ex] 
    \hline\hline \\ [-2.25ex] 
    \textbf{199,468 Unique Bidders} & \textbf{Mean} & \textbf{St. Dev.} & \textbf{Min.} & \textbf{Max.} \\
    \hline \\ [-2.25ex]
    No. Auctions Participated & 4.797 & 15.655 & 1 & 1856 \\ 
    \hline \\ [-2.25ex]
    \textbf{Fraction of Bidders} &  &  &  &  \\ 
    \textbf{Who Won...Auctions} & \textbf{0} & \textbf{$\geq$1} & \textbf{$\geq$2} & \textbf{$\geq$5} \\
    \hline \\ [-2.25ex] 
     & .749 & .251 & .059 & .009 \\ 
    \hline \\ [-1.5ex]
    \multicolumn{5}{l}{$^\dagger$ Calculated for auctions that had at least three bidders.}
\end{tabular}
}
\label{tab:summary-1}
\end{table}

The timing of bids demonstrates regular late bidding, with a significant portion of activity occurring close to the auction’s conclusion. Specifically, 16\% of all bids were submitted in the final five minutes, 34.2\% in the last 15 minutes, 47\% within the last hour, and 63.3\% on the final day. While partly a consequence due to BaT default listing of auctions in ascending order of remaining time, i.e. auctions closest to end are listed first, it also reflects that it is only the most committed bidders that actively participate. Thus providing further support to that bidders are bidding up to their valuation.

The auction platform engaged a total 199,468 unique bidders during our data collection period, each participating in an average of 4.797 auctions. Despite this active engagement, 74.9\% of bidders never won an auction, and only 25.1\% won at least once. Furthermore, only a small fraction of the participants achieved multiple wins: 5.9\% won at least twice, and a mere 0.9\% won five or more auctions.

\section{Empirical Analysis of Online Vehicle Auction Data}\label{sec:discussion} 

We begin our analysis of the trained model by evaluating its predictive accuracy on held-out auction data and assessing its ability to capture key economic relationships in the online vehicle auction market. 
Specifically, we examine the structural estimates produced from the full model $\mathcal{M}^*$ and $\mathcal{H}_2^*$ by validating the recovered valuation and market size distributions against empirical bid distributions. To illustrate the superior performance of our estimation procedure, we provide comparative results with a standard OLS regression and a ``direct'' (joint) estimation of $\mathcal{M}^*$ and $\mathcal{H}_2^*$ (i.e. not training $\mathcal{H}_1$). Finally, we conduct counterfactual simulations to analyze how variations in vehicle descriptions influence estimated demand primitives, allowing us to quantify the economic impact of textual attributes on bidding behavior and price formation. 

\subsection{Prediction Performance of Valuation and Market Size Distribution Estimates}

\begin{figure}
\centering
\includegraphics[width=0.775\textwidth]{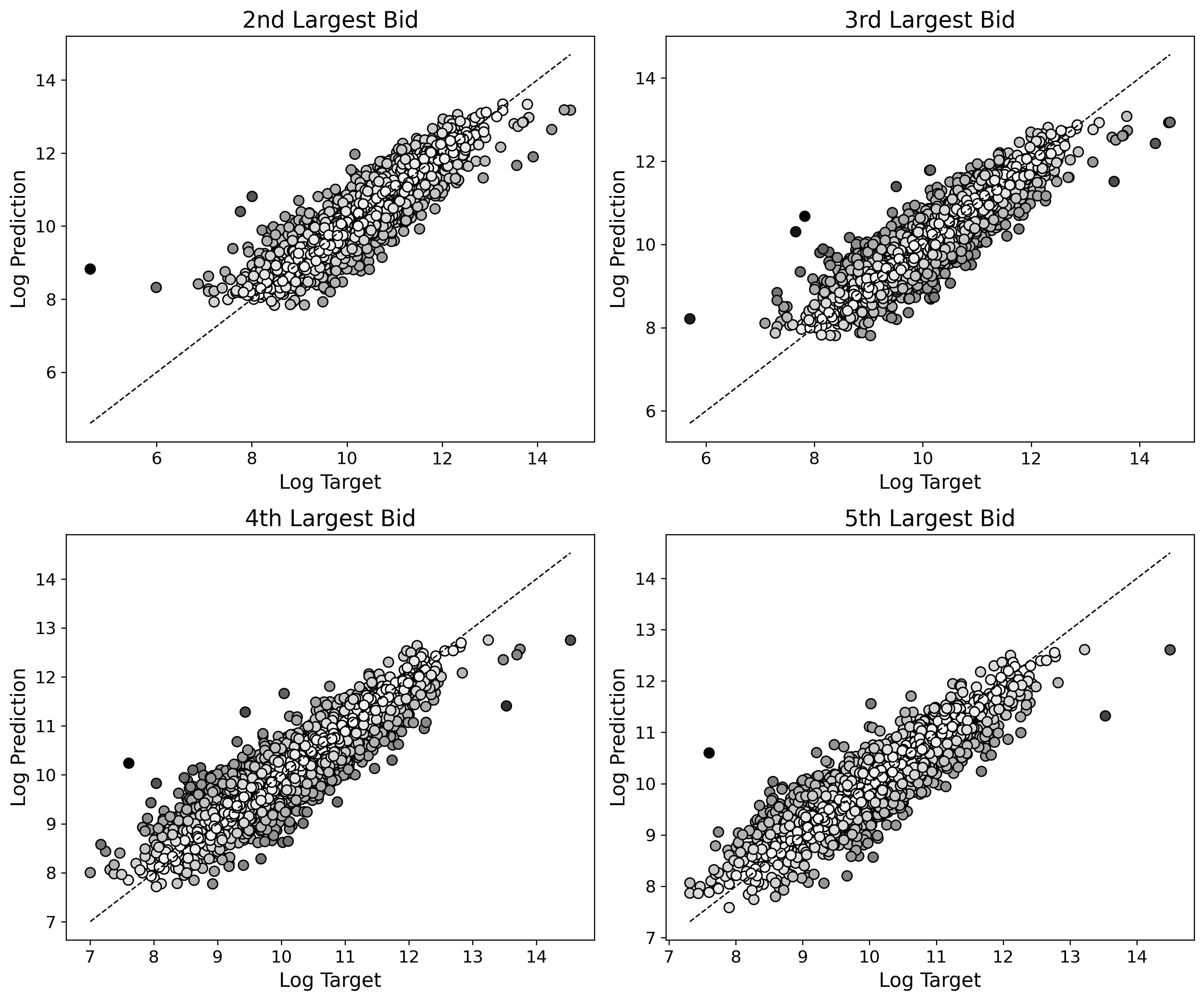}
\caption{Validation Prediction Error on 2$^\text{nd}$, 3$^\text{rd}$, 4$^\text{th}$ and 5$^\text{th}$ Largest Bids (sample size: 5,000)}
\label{fig:stage-2-validation-predictions}
\end{figure}

Figure \ref{fig:stage-2-validation-predictions} presents the nominal prediction performance from the structural two-stage model, evaluated on a validation dataset of 5,000 auctions. Each panel reports the log prediction value for the $2^{nd}$ through $5^{th}$ highest bids against their respective log target values. The predictions were estimated using the inferred valuation and market size distributions. In each panel, observations above and below the 45 degree line indicate instances of over- and under-estimating the (true) bid values, respectively. Overall, we observe the model producing consistent predictions with the observations clustered along the 45 degree line.

Model fit statistics are presented in Table \ref{tab:horserace_results}, which also documents the performance of our two alternative benchmark specifications. This analysis includes a standard OLS estimation of the winning bid, or $2$nd highest bid, and a direct estimation procedure of the attention-based language model where the first stage embedding training of $\mathcal{H}_1$ is ignored, and instead $\mathcal{M}$ and $\mathcal{H}_2$ are jointly estimated. Details of the OLS analysis are presented in Appendix \ref{app:ols-estimation}.

Table~\ref{tab:horserace_results} presents out-of-sample predictive accuracy for the second through fifth-highest bids using the  three estimation procedures. Across all metrics and bid ranks, the two-stage estimator (TS) exhibits consistently superior performance. On the second-highest bid, which serves as the price-setting bid in an English auction, TS achieves a 25.6\% reduction in root mean squared error (RMSE) relative to OLS (.4418 vs. .5940) and a corresponding increase in explanatory power (\(R^2 = 0.779\) for TS vs.\ 0.597 for OLS). The improvements over direct estiamation (DE) are more substantial, with TS reducing RMSE by 54.2\% and increasing \(R^2\) by more than 83 percentage points (from –0.055 to 0.779). These performance differentials persist across higher-order bids, indicating the robustness of the proposed framework.

\begin{table}[htbp]
\centering
\begin{threeparttable}
\caption{Model Fit Metrics for Two-Stage (TS), Direct Estimation (DE), and OLS.}
\label{tab:horserace_results}
\small 
\setlength{\tabcolsep}{5pt} % Slightly adjusted column separation for the new column

% --- MODIFIED TABLE STRUCTURE ---
% The first group now has three 'r' columns (rrr) to accommodate OLS.
\begin{tabular}{@{} l || rrr | rr | rr | rr @{}}
\toprule
% --- MODIFIED TOP HEADER ---
% The "2nd Bid" multicolumn now spans 3 columns.
& \multicolumn{3}{c|}{2nd Bid} & \multicolumn{2}{c|}{3rd Bid} & \multicolumn{2}{c|}{4th Bid} & \multicolumn{2}{c}{5th Bid} \\
\cmidrule(lr){2-4} \cmidrule(lr){5-6} \cmidrule(lr){7-8} \cmidrule(lr){9-10}
% --- MODIFIED SECOND HEADER ---
% "OLS" is added as a model header.
Metric & TS & DE & OLS & TS & DE & TS & DE & TS & DE \\
\midrule
% --- MODIFIED DATA ROWS ---
% The new OLS data is added in the 3rd data column.
RMSE (log)    & \bfseries .4418 & .9649 & .5940 & \bfseries .4271 & .9441 & \bfseries .4166 & .9224 & \bfseries .4167 & .9137 \\
R²            & \bfseries .7789 & -.0550 & .5968 & \bfseries .7899 & -.0267 & \bfseries .7915 & -.0218 & \bfseries .7892 & -.0132 \\
MAPE\%        & \bfseries 37.84 & 141.86 & 49.05 & \bfseries 34.78 & 125.08 & \bfseries 33.45 & 118.03 & \bfseries 33.55 & 114.73 \\
MdAPE\%       & \bfseries 26.06 & 63.21 & 33.73 & \bfseries 26.04 & 60.64 & \bfseries 24.96 & 60.42 & \bfseries 25.77 & 60.71 \\
Hit\%         & \bfseries 20.18 & 8.51 & 17.88 & \bfseries 21.09 & 8.51 & \bfseries 20.66 & 8.09 & \bfseries 20.79 & 8.16 \\
Bias\%        & \bfseries -0.53 & 33.91 & -0.88 & \bfseries -3.44 & 28.74 & \bfseries -3.96 & 26.69 & \bfseries -4.55 & 24.26 \\
\midrule 
% --- MODIFIED N ROW ---
% N is now listed individually for the first group due to the different OLS count.
N             & 4,663 & 4,663 & 4,413 & \multicolumn{2}{c|}{4,172} & \multicolumn{2}{c|}{3,499} & \multicolumn{2}{c}{2,746} \\
\bottomrule
\end{tabular}

% --- MODIFIED NOTES SECTION ---
% A definition for OLS has been added.
\begin{tablenotes}
    \item[\small] \textbf{Note:} %TS: Two-Stage Estimation Procedure; DE: Direct Estimation Procedure; OLS: Ordinary Least Squares. 
    Bold indicates superior performance.
    
    \item[\small] \textbf{Metric Definitions:}
    \begin{itemize} \setlength\itemsep{0.5pt} 
        \item \textbf{MdAPE\%} (Median Absolute Percentage Error): The median of the percentage errors, calculated as $\frac{|\text{Prediction} - \text{Target}|}{|\text{Target}|}$. It represents the typical error and is robust to outliers.
        
        \item \textbf{Hit\%} (Hit Rate): The percentage of predictions where the absolute percentage error is 10\% or less, as defined by the script's `hit\_tolerance` parameter.
        
        \item \textbf{Bias\%} (Multiplicative Bias): Measures systematic over/under-prediction in the original scale. It is calculated from the log-scale errors as $(\exp(\text{mean}(\log(\text{Prediction}) - \log(\text{Target}))) - 1) \times 100$. A value near zero is ideal.
    \end{itemize}
\end{tablenotes}

\end{threeparttable}
\end{table}

Importantly, the two-stage procedure eliminates the substantial upward bias exhibited by the direct estimator. While DE systematically overpredicts all bids, with multiplicative bias ranging from 24.26\% to 33.91\%, the TS model produces mean log prediction errors that correspond to bias levels below 5\% in magnitude and are consistently negative, reflecting mild underprediction. The TS model also outperforms both alternatives on hit rate, defined as the percentage of predictions with less than 10\% absolute percentage error. It achieves hit rates above 20\% across all bid ranks, more than doubling those of DE and modestly outperforming OLS on the second bid (20.18\% vs.\ 17.88\%). In terms of robustness to outliers, the median absolute percentage error (MdAPE) for TS is roughly 26\%, compared to 34\% for OLS and over 60\% for DE.

To leverage the unique economic insights provided by the language model in our estimation procedure, we also report on how each token in a vehcile's description contribute to the value of the 2$^\text{nd}$-largest (i.e. price-setting) bid. To extract the ways textual descriptions influence bid predictions we apply Integrated Gradients (IG; \citealp{sundararajan2017axiomatic}), which is an interpretability method that decomposes continuous prediction values into token-level contributions. IG quantifies each token's incremental contribution to the model's prediction relative to a neutral baseline embedding (we employ the \texttt{[PAD]} token as this baseline). Formally, IG integrates the gradient of the model's bid prediction function along a straight-line interpolation between the baseline and actual input embeddings, using, in our case, $K = 256$ discrete intervals. The resulting attribution scores fulfill a completeness criterion such that their sum closely approximates the overall prediction difference within a controlled tolerance of $\max\left(10^{-3},\, 0.02\cdot |f(x)-f(x')|\right)$. Token-level attributions are further aggregated to meaningful word-level insights, excluding special tokens that lack interpretative significance. 

\begin{figure}[!htbp]
\centering 
\begin{tikzpicture}
\begin{axis}[
    hide axis,
    scale only axis,
    height=0pt,
    width=10cm,
    colorbar horizontal,
    point meta min=-1, 
    point meta max= 1,
    colorbar style={
        width=10cm,
        height=0.5cm,
        xtick={-1,-0.5,0,0.5,1},
        xticklabels={$-1$,$-0.5$,$0$,$0.5$,$1$},
        xticklabel style={text height=1.5ex},
        xtick style={draw=none},
        colormap name=divRedWhiteBlue
    },
]
    % dummy invisible plot just to force the bar to draw
    \addplot [draw=none] coordinates {(-1,0) (1,0)};
\end{axis}
\end{tikzpicture}

\raggedright 
\input{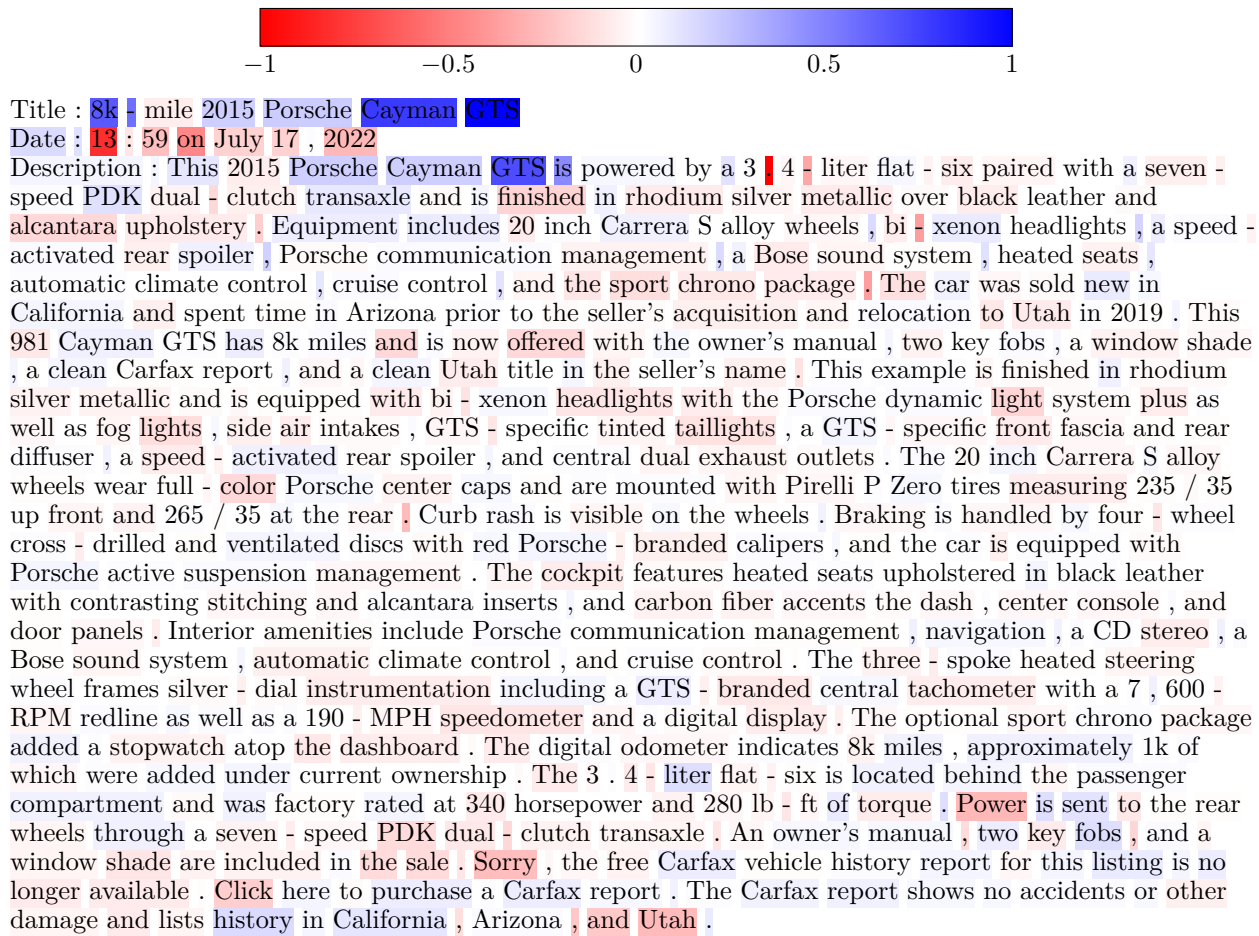}
\caption{Stage 1 Estimation: Normalized Token Contributions to Price-setting Bid}
\label{fig:stage-1-values-example}
\end{figure}

Figure~\ref{fig:stage-1-values-example} provides a visualization of normalized token attributions aggregated at the word-level for an illustrative auction listing. Positive and negative influences on predicted log-bid values are color-coded in shades of blue and red, respectively, enabling straightforward identification of description elements driving bidding outcomes. This granular interpretability highlights specific textual attributes, such as vehicle condition, rarity, or desirable features, that critically influence the marginal bid value. Insights derived from this stage of analysis not only enrich theoretical understanding of hedonic pricing mechanisms but also offer practical guidance for sellers aiming to optimize listing descriptions for enhanced market outcomes. Figure~\ref{fig:stage-1-views-example} in Appendix~\ref{app:additional-analyses} illustrates an example of the words driving the number of potential bidders to the auction.

\subsection{Zero-Shot Learning: The DMC DeLorean}

A natural test of the Stage 2 model's ability to generalize beyond its training distribution is to evaluate its predictive performance on vehicles that were entirely absent from the training set. To assess this capability, we deliberately withheld a specific and unusual make and model from both the training and validation datasets: the DMC DeLorean. The vehicle, famous from the \emph{Back to the Future} movie franchise, possesses a distinct combination of design elements, historical significance, and cultural associations that are unlikely to be collectively shared by any other vehicles in the dataset. Consequently, accurate predictions on this category of vehicles would provide strong evidence that the model has successfully learned generalizable demand patterns within the target market.

\begin{figure}
\centering
\includegraphics[width=0.775\textwidth]{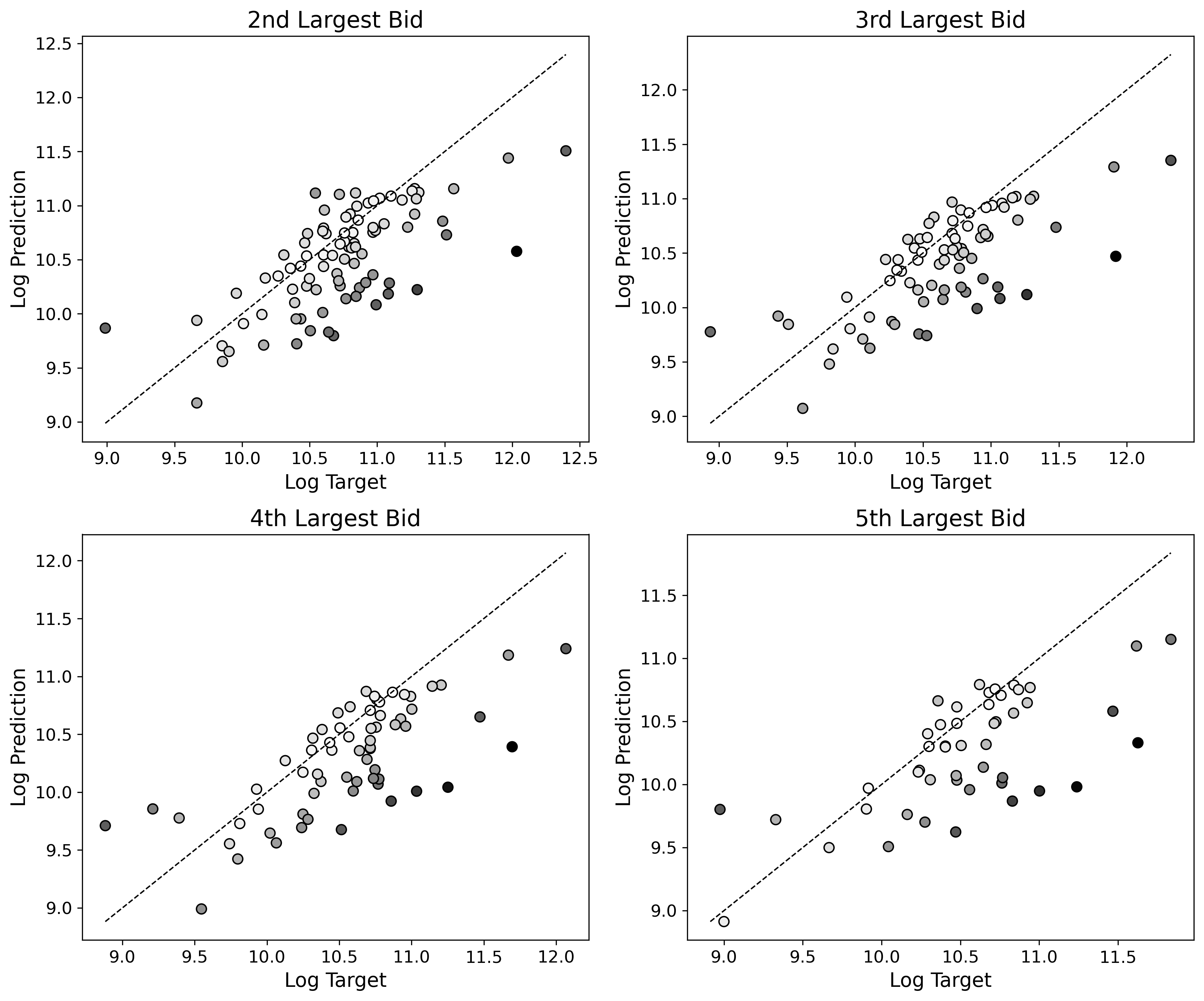}
\caption{Zero-shot Predictions: DMC DeLorean (sample size: 93)}
\label{fig:delorean-validations-1}
\end{figure}

In total, the dataset contained 278 bid observations from 93 auctions for DMC DeLoreans. Figure \ref{fig:delorean-validations-1} presents the nominal prediction errors on the second-to-fifth largest bids observed in these zero-shot auctions. The prediction errors for the zero-shot evaluation of the DMC DeLorean exhibit distinct characteristics relative to the broader validation set. The model maintains relatively stable predictive accuracy in terms of log-scale error dispersion, with RMSLE values ranging from 0.44 to 0.50 across the second- through fifth-highest bids. Nevertheless, we do observe a systematic bias across all bid ranks where the model consistently underestimates realized prices. This persistent undervaluation likely arises because the model does not, and indeed, given no relevant training data, cannot, fully capture the substantial and idiosyncratic premium placed on the DeLorean, which partly may be due to its iconic association with \textit{Back to the Future}. The vehicle's cultural significance, rather than conventional automotive attributes captured by structured auction data, appears to drive much of its market value. 

Our fine-tuned model's ability to make robust out-of-sample predictions on vehicles it has never encountered directly stems from the deep representations learned during pretraining \citep{wei2021finetuned}. Large-scale language models, such as mGTE, are trained on extensive corpora that include information about a wide range of consumer goods, historical events, and market trends. It is likely that the foundational model trained in \cite{li2023gte} was exposed to content about the DeLorean, including design and production history, collector status, and pop culture significance. During the fine-tuning process in Stage 1, some of these latent representations were likely preserved and refined within the context of auction markets, allowing the model to infer a reasonable valuation for the DeLorean by positioning it relative to other vehicles with similar attributes in the dataset. However, this explanation remains speculative, as testing whether and how these representations were preserved is beyond the scope of this paper. Future research could more directly investigate the extent to which pretrained language models retain and adapt knowledge representations during fine-tuning, particularly in economic applications \citep[see][]{templeton2024scaling}. 

\subsection{Counterfactual Simulations: Systematically Varying Vehicle Mileage} 

To assess the economic consistency of our trained model, we conduct a series of counterfactual simulations that systematically vary a key determinant of vehicle value: mileage. Vehicle mileage is one of the most robust indicators of depreciation, influencing resale values across all makes and models. Therefore, as a robustness check, we test whether the model's price-setting bid predictions exhibit an economically reasonable response to changes in reported mileage. Due to the computational cost of this analysis, we randomly select a subset of 1,000 vehicles from the validation dataset and use a large language model (LLAMA 3.1 70B) to generate modified versions of each vehicle's description while holding all other details constant. Specifically, we replace the mileage value in the original description with one of six predetermined values: 25k, 50k, 75k, 100k, 125k, and 150k miles. Each edited description is then processed by the trained Stage 2 model to generate a new set of predicted price-setting bids.

\begin{figure}
\centering
\includegraphics[width=0.95\textwidth]{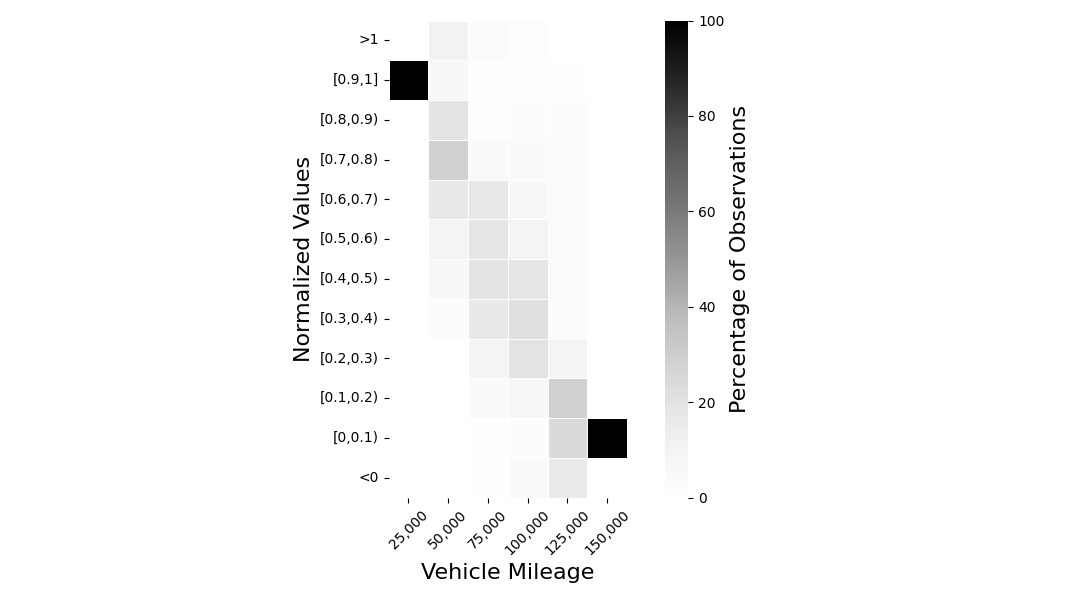}
\caption{Counterfactual Mileage Simulations (sample size: 1,000)}
\label{fig:mileage-counterfactuals-1}
\end{figure}

Figure \ref{fig:mileage-counterfactuals-1} presents the results of this experiment, where the price-setting bid predictions are normalized to 1 for the 25k-mile counterfactual and 0 for the 150k-mile counterfactual. Across the vehicles included in the analysis, the model predicts a strong monotonic decline in the price-setting bid as mileage increases. This finding aligns with standard economic expectations regarding vehicle depreciation, and reinforcing the model's ability to appropriately internalize the role of accumulated mileage in determining vehicle value.

These results also highlight the utility of our framework for evaluating the marginal effects of key vehicle attributes on expected auction outcomes. By systematically modifying vehicle descriptions and observing the model's predictions, researchers and practitioners can conduct detailed counterfactual analyses that would be infeasible to observe in real-world auction data. Future work could extend this approach by examining other vehicle characteristics, such as accident history, number of previous owners, or specific aftermarket modifications, to further explore the model’s sensitivity to economically relevant factors. 

\section{Conclusion} \label{sec:conclusion} 

This paper introduces a novel approach to demand estimation that integrates language models with structural econometric techniques to infer primitives of demand from unstructured textual descriptions. By training an encoder-only mGTE model in a two-stage process, we demonstrate how natural language descriptions of auctioned vehicles can be systematically transformed into structured economic representations that capture latent demand primitives. Our empirical validation using online vehicle auction data confirms that the proposed model effectively encodes market-relevant information, yielding competitive predictive accuracy on auction outcomes while preserving interpretability through its structural estimation framework.

A key strength of our approach is its ability to generalize beyond the training distribution, as evidenced by the model’s performance on the validation dataset and the zero-shot predictions on the DMC DeLorean. This result highlights the potential of language models to leverage their pre-trained knowledge when making economic inferences in unfamiliar contexts, a property that could be further explored in other market settings. Additionally, our counterfactual simulations on vehicle mileage confirm the model’s ability to produce economically coherent predictions, reinforcing its capacity to extract meaningful demand signals from textual descriptions.

These findings have several important implications for both researchers and practitioners. For firms and digital platforms, the ability to extract consumer demand signals directly from product descriptions offers a new avenue for designing ``smart markets'' via strategic pricing and inventory management \citep{bichler2010research}. Rather than relying solely on historical sales data, firms can leverage language models to predict demand for new products, optimize product descriptions for maximum market appeal, and conduct counterfactual analysis to explore pricing and positioning strategies. From a methodological perspective, our study contributes to the growing literature on machine learning applications in economics by demonstrating how deep learning models can be integrated with structural estimation techniques to improve demand inference.

Despite these contributions, important avenues for future research remain. First, while our approach successfully captures demand primitives from textual data, further work is needed to understand how fine-tuned language models retain and adapt pre-trained knowledge for economic inference. The extent to which language models internalize and preserve economic reasoning during domain-specific training is an open question that warrants further exploration using interpretability methods, such as feature attribution and causal probing techniques. Second, while our empirical application focuses on English auctions, the underlying framework could be extended to other market settings, including posted-price markets, bargaining environments, and two-sided platforms where product descriptions and buyer preferences interact dynamically. Lastly, given the rapid evolution of language model architectures, future studies could investigate how different model families (e.g., multimodal models with vision capabilities) compare in their ability to encode and generalize economic information.

\section*{Acknowledgments}
None. 

%Bibliography
\bibliographystyle{unsrt}  
\bibliography{refs.bib}  

\begin{thebibliography}{10}

\bibitem{rosen1974hedonic}
Sherwin Rosen.
\newblock Hedonic prices and implicit markets: product differentiation in pure competition.
\newblock {\em Journal of Political Economy}, 82(1):34--55, 1974.

\bibitem{ludwig2024machine}
Jens Ludwig and Sendhil Mullainathan.
\newblock Machine learning as a tool for hypothesis generation.
\newblock {\em Quarterly Journal of Economics}, 139(2):751--827, 2024.

\bibitem{cropper1988choice}
Maureen~L Cropper, Leland~B Deck, and Kenenth~E McConnell.
\newblock On the choice of funtional form for hedonic price functions.
\newblock {\em Review of Economics and Statistics}, pages 668--675, 1988.

\bibitem{mikolov2013linguistic}
Tom{\'a}{\v{s}} Mikolov, Wen-tau Yih, and Geoffrey Zweig.
\newblock Linguistic regularities in continuous space word representations.
\newblock In {\em Proceedings of the 2013 conference of the north american chapter of the association for computational linguistics: Human language technologies}, pages 746--751, 2013.

\bibitem{qiu2020pre}
Xipeng Qiu, Tianxiang Sun, Yige Xu, Yunfan Shao, Ning Dai, and Xuanjing Huang.
\newblock Pre-trained models for natural language processing: A survey.
\newblock {\em Science China Technological Sciences}, 63(10):1872--1897, 2020.

\bibitem{devlin2018bert}
Jacob Devlin, Ming-Wei Chang, Kenton Lee, and Kristina Toutanova.
\newblock Bert: Pre-training of deep bidirectional transformers for language understanding.
\newblock {\em arXiv preprint arXiv:1810.04805}, 2018.

\bibitem{athey2018impact}
Susan Athey.
\newblock The impact of machine learning on economics.
\newblock In {\em The economics of artificial intelligence: An agenda}, pages 507--547. University of Chicago Press, 2018.

\bibitem{mehta2021sell}
Sameer Mehta, Milind Dawande, Ganesh Janakiraman, and Vijay Mookerjee.
\newblock How to sell a data set? pricing policies for data monetization.
\newblock {\em Information Systems Research}, 32(4):1281--1297, 2021.

\bibitem{van2022overcoming}
Benjamin Van~Giffen, Dennis Herhausen, and Tobias Fahse.
\newblock Overcoming the pitfalls and perils of algorithms: A classification of machine learning biases and mitigation methods.
\newblock {\em Journal of Business Research}, 144:93--106, 2022.

\bibitem{abbasi2024pathways}
Ahmed Abbasi, Jeffrey Parsons, Gautam Pant, Olivia R~Liu Sheng, and Suprateek Sarker.
\newblock Pathways for design research on artificial intelligence.
\newblock {\em Information Systems Research}, 35(2):441--459, 2024.

\bibitem{sarker2025advancing}
Suprateek Sarker, Hillol Bala, Yili Hong, Atreyi Kankanhalli, Matti Rossi, Bin Gu, and Gal Oestreicher-Singer.
\newblock Advancing next-generation multimethod research in information systems: A framework and some recommendations for authors and evaluators.
\newblock {\em Information Systems Research}, 36(2):647--668, 2025.

\bibitem{vaswani2017attention}
Ashish Vaswani, Noam Shazeer, Niki Parmar, Jakob Uszkoreit, Llion Jones, Aidan~N Gomez, {\L}ukasz Kaiser, and Illia Polosukhin.
\newblock Attention is all you need.
\newblock {\em Advances in Neural Information Processing Systems}, 30, 2017.

\bibitem{radford2019language}
Alec Radford, Jeffrey Wu, Rewon Child, David Luan, Dario Amodei, Ilya Sutskever, et~al.
\newblock Language models are unsupervised multitask learners.
\newblock {\em OpenAI blog}, 1(8):9, 2019.

\bibitem{jumper2021highly}
John Jumper, Richard Evans, Alexander Pritzel, Tim Green, Michael Figurnov, Olaf Ronneberger, Kathryn Tunyasuvunakool, Russ Bates, Augustin {\v{Z}}{\'\i}dek, Anna Potapenko, et~al.
\newblock Highly accurate protein structure prediction with alphafold.
\newblock {\em Nature}, 596(7873):583--589, 2021.

\bibitem{chen2021transunet}
Jieneng Chen, Yongyi Lu, Qihang Yu, Xiangde Luo, Ehsan Adeli, Yan Wang, Le~Lu, Alan~L Yuille, and Yuyin Zhou.
\newblock Transunet: Transformers make strong encoders for medical image segmentation.
\newblock {\em arXiv preprint arXiv:2102.04306}, 2021.

\bibitem{Chen2024}
Zenan Chen and Jason Chan.
\newblock Large language model in creative work: The role of collaboration modality and user expertise.
\newblock {\em Management Science}, 70(12):3450--3472, 2024.

\bibitem{lin2024automated}
Fangyu Lin, Sagar Samtani, Hongyi Zhu, Brandimarte Laura, and Hsinchun Chen.
\newblock Automated analysis of changes in privacy policies: A structured self-attentive sentence embedding approach.
\newblock {\em MIS Quarterly}, 48(4), 2024.

\bibitem{zhou2024can}
Yaxian Zhou, Yufei Yuan, Kai Huang, and Xiangpei Hu.
\newblock Can chatgpt perform a grounded theory approach to do risk analysis? an empirical study.
\newblock {\em Journal of Management Information Systems}, 41(4):982--1015, 2024.

\bibitem{wang2025predicting}
Wen Wang, Mi~Zhou, Beibei Li, and Honglei Zhuang.
\newblock Predicting instructor performance in online education: An interpretable hierarchical transformer with contextual attention.
\newblock {\em Information Systems Research}, 2025.

\bibitem{gentzkow2019text}
Matthew Gentzkow, Bryan Kelly, and Matt Taddy.
\newblock Text as data.
\newblock {\em Journal of Economic Literature}, 57(3):535--574, 2019.

\bibitem{Berger2020}
Jonah Berger, Ashlee Humphreys, Stefan Ludwig, Wendy~W. Moe, Oded Netzer, and David~A. Schweidel.
\newblock Uniting the tribes: Using text for marketing insight.
\newblock {\em Journal of Marketing}, 84(1):1--25, 2020.

\bibitem{ash2023text}
Elliott Ash and Stephen Hansen.
\newblock Text algorithms in economics.
\newblock {\em Annual Review of Economics}, 15:659--688, 2023.

\bibitem{deKok2025}
Ties de~Kok.
\newblock Chatgpt for textual analysis? how to use generative llms in accounting research.
\newblock {\em Management Science}, 71(1):123--145, 2025.

\bibitem{Puranam2021}
Dinesh Puranam, Vrinda Kadiyali, and Vithala Narayan.
\newblock The impact of increase in minimum wages on consumer perceptions of service: A transformer model of online restaurant reviews.
\newblock {\em Marketing Science}, 40(5):985--1004, 2021.

\bibitem{yang2023getting}
Kai Yang, Raymond~YK Lau, and Ahmed Abbasi.
\newblock Getting personal: A deep learning artifact for text-based measurement of personality.
\newblock {\em Information Systems Research}, 34(1):194--222, 2023.

\bibitem{ma2023beyond}
Liye Ma and Lan Luo.
\newblock Beyond fake or genuine--the effect of large language models (llms) on the content and sentiment of product reviews.
\newblock {\em USC Marshall School of Business Research Paper Sponsored by iORB, No. Forthcoming}, 2023.

\bibitem{qiu2023consumer}
Liying Qiu, Param~Vir Singh, and Kannan Srinivasan.
\newblock Consumer risk preferences elicitation from large language models.
\newblock {\em Available at SSRN 4526072}, 2023.

\bibitem{mccarthy2024fin}
Shawn McCarthy and Gita Alaghband.
\newblock Fin-alice: Artificial linguistic intelligence causal econometrics.
\newblock {\em Journal of Risk and Financial Management}, 17(12):537, 2024.

\bibitem{Timoshenko2019}
Anna Timoshenko and John~R. Hauser.
\newblock Identifying customer needs from user-generated content.
\newblock {\em Marketing Science}, 38(1):1--20, 2019.

\bibitem{Liu2023}
Xiao Liu.
\newblock Deep learning in marketing: a review and research agenda.
\newblock {\em Artificial Intelligence in Marketing}, pages 239--271, 2023.

\bibitem{wang2024large}
Mengxin Wang, Dennis~J Zhang, and Heng Zhang.
\newblock Large language models for market research: A data-augmentation approach.
\newblock {\em arXiv preprint arXiv:2412.19363}, 2024.

\bibitem{chen2025conversation}
Yanzhen Chen, Huaxia Rui, and Andrew~B Whinston.
\newblock Conversation analytics: Can machines read between the lines in real-time strategic conversations?
\newblock {\em Information Systems Research}, 36(1):440--455, 2025.

\bibitem{athey2019machine}
Susan Athey and Guido~W Imbens.
\newblock Machine learning methods that economists should know about.
\newblock {\em Annual Review of Economics}, 11(1):685--725, 2019.

\bibitem{adam2024machine}
Hammaad Adam, Pu~He, and Fanyin Zheng.
\newblock Machine learning for demand estimation in long tail markets.
\newblock {\em Management Science}, 70(8):5040--5065, 2024.

\bibitem{aceves2024mobilizing}
Pedro Aceves and James~A Evans.
\newblock Mobilizing conceptual spaces: How word embedding models can inform measurement and theory within organization science.
\newblock {\em Organization Science}, 35(3):788--814, 2024.

\bibitem{dillion2023can}
Danica Dillion, Niket Tandon, Yuling Gu, and Kurt Gray.
\newblock Can ai language models replace human participants?
\newblock {\em Trends in Cognitive Sciences}, 27(7):597--600, 2023.

\bibitem{wu2023large}
Patrick~Y Wu, Jonathan Nagler, Joshua~A Tucker, and Solomon Messing.
\newblock Large language models can be used to estimate the latent positions of politicians.
\newblock {\em arXiv preprint arXiv:2303.12057}, 2023.

\bibitem{goli2024frontiers}
Ali Goli and Amandeep Singh.
\newblock Frontiers: Can large language models capture human preferences?
\newblock {\em Marketing Science}, 43(4):709--722, 2024.

\bibitem{zhu2024language}
Jian-Qiao Zhu, Haijiang Yan, and Thomas~L Griffiths.
\newblock Language models trained to do arithmetic predict human risky and intertemporal choice.
\newblock {\em arXiv preprint arXiv:2405.19313}, 2024.

\bibitem{olah2020zoom}
Chris Olah, Nick Cammarata, Ludwig Schubert, Gabriel Goh, Michael Petrov, and Shan Carter.
\newblock Zoom in: An introduction to circuits.
\newblock {\em Distill}, 5(3):e00024--001, 2020.

\bibitem{elhage2022toy}
Nelson Elhage, Tristan Hume, Catherine Olsson, Nicholas Schiefer, Tom Henighan, Shauna Kravec, Zac Hatfield-Dodds, Robert Lasenby, Dawn Drain, Carol Chen, et~al.
\newblock Toy models of superposition.
\newblock {\em arXiv preprint arXiv:2209.10652}, 2022.

\bibitem{templeton2024scaling}
Adly Templeton.
\newblock {\em Scaling monosemanticity: Extracting interpretable features from claude 3 sonnet}.
\newblock Anthropic, 2024.

\bibitem{verma2024counterfactual}
Sahil Verma, Varich Boonsanong, Minh Hoang, Keegan Hines, John Dickerson, and Chirag Shah.
\newblock Counterfactual explanations and algorithmic recourses for machine learning: A review.
\newblock {\em ACM Computing Surveys}, 56(12):1--42, 2024.

\bibitem{wei2021finetuned}
Jason Wei, Maarten Bosma, Vincent~Y Zhao, Kelvin Guu, Adams~Wei Yu, Brian Lester, Nan Du, Andrew~M Dai, and Quoc~V Le.
\newblock Finetuned language models are zero-shot learners.
\newblock {\em arXiv preprint arXiv:2109.01652}, 2021.

\bibitem{vafa2024estimating}
Keyon Vafa, Susan Athey, and David~M Blei.
\newblock Estimating wage disparities using foundation models.
\newblock {\em arXiv preprint arXiv:2409.09894}, 2024.

\bibitem{li2023gte}
Zehan Li, Xin Zhang, Yanzhao Zhang, Dingkun Long, Pengjun Xie, and Meishan Zhang.
\newblock Towards general text embeddings with multi-stage contrastive learning, 2023.

\bibitem{turian2010word}
Joseph Turian, Lev Ratinov, and Yoshua Bengio.
\newblock Word representations: a simple and general method for semi-supervised learning.
\newblock In {\em Proceedings of the 48th annual meeting of the association for computational linguistics}, pages 384--394, 2010.

\bibitem{vylomova2015take}
Ekaterina Vylomova, Laura Rimell, Trevor Cohn, and Timothy Baldwin.
\newblock Take and took, gaggle and goose, book and read: Evaluating the utility of vector differences for lexical relation learning.
\newblock {\em arXiv preprint arXiv:1509.01692}, 2015.

\bibitem{li2020sentence}
Bohan Li, Hao Zhou, Junxian He, Mingxuan Wang, Yiming Yang, and Lei Li.
\newblock On the sentence embeddings from pre-trained language models.
\newblock {\em arXiv preprint arXiv:2011.05864}, 2020.

\bibitem{mullainathan2017machine}
Sendhil Mullainathan and Jann Spiess.
\newblock Machine learning: an applied econometric approach.
\newblock {\em Journal of Economic Perspectives}, 31(2):87--106, 2017.

\bibitem{iskhakov2020machine}
Fedor Iskhakov, John Rust, and Bertel Schjerning.
\newblock Machine learning and structural econometrics: contrasts and synergies.
\newblock {\em The Econometrics Journal}, 23(3):S81--S124, 2020.

\bibitem{gallant1987semi}
A~Ronald Gallant and Douglas~W Nychka.
\newblock Semi-nonparametric maximum likelihood estimation.
\newblock {\em Econometrica}, pages 363--390, 1987.

\bibitem{fenton1996convergence}
Victor~M Fenton and A~Ronald Gallant.
\newblock Convergence rates of snp density estimators.
\newblock {\em Econometrica}, pages 719--727, 1996.

\bibitem{fenton1996qualitative}
Victor~M Fenton and A~Ronald Gallant.
\newblock Qualitative and asymptotic performance of snp density estimators.
\newblock {\em Journal of Econometrics}, 74(1):77--118, 1996.

\bibitem{farrell2020deep}
Max~H Farrell, Tengyuan Liang, and Sanjog Misra.
\newblock Deep learning for individual heterogeneity: An automatic inference framework.
\newblock {\em arXiv preprint arXiv:2010.14694}, 2020.

\bibitem{wolfstetter1996auctions}
Elmar Wolfstetter.
\newblock Auctions: an introduction.
\newblock {\em Journal of Economic Surveys}, 10(4):367--420, 1996.

\bibitem{lu2019information}
Yixin Lu, Alok Gupta, Wolfgang Ketter, and Eric Van~Heck.
\newblock Information transparency in business-to-business auction markets: The role of winner identity disclosure.
\newblock {\em Management Science}, 65(9):4261--4279, 2019.

\bibitem{avery1998strategic}
Christopher Avery.
\newblock Strategic jump bidding in english auctions.
\newblock {\em The Review of Economic Studies}, 65(2):185--210, 1998.

\bibitem{easley2004jump}
Robert~F Easley and Rafael Tenorio.
\newblock Jump bidding strategies in internet auctions.
\newblock {\em Management Science}, 50(10):1407--1419, 2004.

\bibitem{liu2020preserving}
De~Liu and Adib Bagh.
\newblock Preserving bidder privacy in assignment auctions: design and measurement.
\newblock {\em Management Science}, 66(7):3162--3182, 2020.

\bibitem{morokoff1995quasi}
William~J Morokoff and Russel~E Caflisch.
\newblock Quasi-monte carlo integration.
\newblock {\em Journal of Computational Physics}, 122(2):218--230, 1995.

\bibitem{jackel2002monte}
Peter J{\"a}ckel.
\newblock {\em Monte Carlo methods in finance}, volume~5.
\newblock John Wiley \& Sons, 2002.

\bibitem{sundararajan2017axiomatic}
Mukund Sundararajan, Ankur Taly, and Qiqi Yan.
\newblock Axiomatic attribution for deep networks.
\newblock In {\em International conference on machine learning}, pages 3319--3328. PMLR, 2017.

\bibitem{bichler2010research}
Martin Bichler, Alok Gupta, and Wolfgang Ketter.
\newblock Research commentary—designing smart markets.
\newblock {\em Information Systems Research}, 21(4):688--699, 2010.

\bibitem{yun2019transformers}
Chulhee Yun, Srinadh Bhojanapalli, Ankit~Singh Rawat, Sashank~J Reddi, and Sanjiv Kumar.
\newblock Are transformers universal approximators of sequence-to-sequence functions?
\newblock {\em arXiv preprint arXiv:1912.10077}, 2019.

\bibitem{lu2020universal}
Yulong Lu and Jianfeng Lu.
\newblock A universal approximation theorem of deep neural networks for expressing probability distributions.
\newblock {\em Advances in Neural Information Processing Systems}, 33:3094--3105, 2020.

\bibitem{hornik1989multilayer}
Kurt Hornik, Maxwell Stinchcombe, and Halbert White.
\newblock Multilayer feedforward networks are universal approximators.
\newblock {\em Neural Networks}, 2(5):359--366, 1989.

\bibitem{pakes2003reconsideration}
Ariel Pakes.
\newblock A reconsideration of hedonic price indexes with an application to pc’s.
\newblock {\em American Economic Review}, 93(5):1578--1596, 2003.

\end{thebibliography}

\appendix

\section{Proofs of Propositions}\label{app:proofs}

\subsection{Proposition 1} 

Let $d$ denote a text description and $\vec{y} \in \mathbb{R}^m$ represent the vector of $m$ economic outcomes associated with $d$. Let $\mathcal{M}$ be a language model that maps a text description $d$ to an embedding vector $\{\vec{e}_0\} \in \mathbb{R}^q$. Let $\mathcal{H}_1$ be a projection function that maps these embeddings to predicted economic outcomes $\hat{\vec{y}} \in \mathbb{R}^m$. The composition of these functions gives us $\hat{\vec{y}} = \mathcal{H}_1 \circ \mathcal{M}(d)$.

To establish the claim, we must show that for any error tolerance $\epsilon > 0$, there exist parameterized versions of these models $\mathcal{M}^*$ and $\mathcal{H}_1^*$ such that
\begin{equation*}
    \|y - \mathcal{H}_{1}^{*}\circ \mathcal{M}^{*}(d)\|_2 < \epsilon
\end{equation*}
for all $(d,y)$ pairs in our dataset $D$.

By the Universal Approximation Theorem for Transformers \citep{yun2019transformers}\footnote{See also \cite{lu2020universal}.}, for any continuous function $f$ mapping sequences to sequences and any $\epsilon > 0$, there exists a transformer network $T$ such that
\begin{equation*}
    \|f(x) - T(x)\|_2 < \epsilon/2
\end{equation*}
for all inputs $x$ in the domain. Similarly, by the classical Universal Approximation Theorem for Feed-Forward Networks \citep{hornik1989multilayer}, for any continuous function $g$ and any $\epsilon > 0$, there exists a feed-forward network $H$ such that
\begin{equation*}
    \|g(\vec{e}) - H(\vec{e})\|_2 < \epsilon/2
\end{equation*}
for all embeddings $\vec{e}$ in the relevant domain. Let $f^*$ be an ideal function that maps text descriptions directly to embeddings that perfectly encode the economic outcomes, and $g^*$ be a function that maps these ideal embeddings to the true economic outcomes. By the first theorem, there exists a language model $\mathcal{M}^{*}$ such that
\begin{equation*}
    \|f^*(d) - \mathcal{M}^{*}(d)\|_2 < \delta
\end{equation*}
where $\delta$ is chosen to ensure that the subsequent projection error is sufficiently small. By the second theorem, there exists a projection function $\mathcal{H}_{1}^{*}$ such that 
\begin{equation*}
    \|g^*(\vec{e}) - \mathcal{H}_{1}^{*}(\vec{e})\|_2 < \epsilon/2
\end{equation*}
for all embeddings $\vec{e}$ in the range of $\mathcal{M}^{*}$. Assuming $\mathcal{H}_{1}^{*}$ is $L$-Lipschitz continuous, 
\begin{equation*}
    \|\mathcal{H}_{1}^{*}\circ f^*(d) - \mathcal{H}_{1}^{*}\circ \mathcal{M}^{*}(d)\|_2 \leq L \cdot \|f^*(d) - \mathcal{M}^{*}(d)\|_2 < L \cdot \delta.
\end{equation*}
We can choose $\delta = \epsilon/(2L)$ to ensure $L \cdot \delta < \epsilon/2$. Combining these inequalities through the triangle inequality, 
\begin{equation*}
\begin{split}
\|y - \mathcal{H}_{1}^{*}\circ \mathcal{M}^{*}(d)\|_2 &= \|g^*(f^*(d)) - \mathcal{H}_{1}^{*}\circ \mathcal{M}^{*}(d)\|_2 \\
&\leq \|g^*(f^*(d)) - \mathcal{H}_{1}^{*}(f^*(d))\|_2 + \|\mathcal{H}_{1}^{*}\circ f^*(d) - \mathcal{H}_{1}^{*}\circ \mathcal{M}^{*}(d)\|_2 \\
&< \epsilon/2 + L \cdot \delta \\
&< \epsilon/2 + \epsilon/2 \\
&= \epsilon
\end{split} 
\end{equation*}
Therefore, the composition $\mathcal{H}_{1}^{*} \circ \mathcal{M}^{*}$ can approximate the mapping of text descriptions to economic outcomes with arbitrary precision $\epsilon$. To establish that economic information can be recovered from the embeddings, we note that $\mathcal{H}_{1}^{*}$ serves as a decoding function. The embeddings $\{\vec{e}_j\}$ produced by $\mathcal{M}_{1}^{*}$ must contain all the information needed to reconstruct $y$ within the specified error bound, otherwise $\mathcal{H}_{1}^{*}\circ \mathcal{M}^{*}(d)$ could not achieve the arbitrary precision approximation we proved above. 

\section{Linear Hedonic Estimation (OLS)} \label{app:ols-estimation}

We estimate a linear hedonic pricing model for comparison purposes to our attention-based language model. Our empirical objective is to estimate the expected second-highest bid, conditional on physical and categorical features of the car, by fitting a linear model of the form:
\begin{align}
\log(\text{Price}_i) &= \beta_0 + \beta_1 \log(\text{Mileage}_i) + \beta_2 \log(\text{Horsepower}_i) + \beta_3 \text{Age}_i \nonumber \\
&\quad + \beta_4 \text{TimeTrend}_i + \beta_5 \text{Automatic}_i + \gamma' \text{EngineType}_i + \theta' \text{Year}_i + \delta' \text{Cohort}_i + \varepsilon_i
\end{align}

\noindent where $\log(\text{Price}_i)$ is the log of the final auction price for vehicle $i$, and the covariates describe vehicle condition, performance characteristics, market timing, and cohort identity.

The model is estimated on a dataset of 17,695 random auction records from our original dataset (we withhold 4,413, or 20\%, for our validation tests). All vehicles included have non-missing values for key variables and pass standard validity checks (e.g., positive mileage and horsepower). The feature engineering process transforms key continuous variables (mileage, horsepower) using logarithmic transformations to stabilize variance and reflect diminishing marginal valuation. Age is computed as the difference between the sale year and the vehicle's model year. Transmission type is represented as a binary indicator for automatic gearboxes. Cohort fixed effects are constructed at the make-model level for high-frequency observations and at the make level otherwise. To capture secular changes in demand, we include both a normalized time trend and annual dummy variables. Engine configuration is controlled for using cylinder-count dummies, with two-cylinder engines serving as the reference category.

We estimate the model using pooled ordinary least squares (OLS) and report robust standard errors. The dependent variable is the log of the second largest bid, which ensures that coefficient estimates can be interpreted approximately as percentage changes. This specification is widely used in the hedonic pricing literature, particularly in automotive and real estate contexts, where both physical features and brand-specific reputations jointly shape willingness-to-pay \citep{rosen1974hedonic, pakes2003reconsideration}. By incorporating fixed effects for make/model and year, the specification accounts for both product-level unobservables and market-wide temporal shifts.

\begin{table}[htbp]
\centering
\begin{threeparttable}
\caption{OLS Hedonic Price Model Parameter Estimates (Training Set)}
\label{tab:hedonic_parameters}
\small
\setlength{\tabcolsep}{6pt}

\begin{tabular}{@{} l c c @{}}
\toprule
Variable & Coefficient & Std. Error \\
\midrule
% --- Main Variables ---
Intercept & 8.403*** & (0.436) \\
Log(Mileage) & -0.077*** & (0.002) \\
Log(Horsepower) & 0.283*** & (0.010) \\
Vehicle Age & -0.001** & (0.0005) \\
Time Trend & 0.102 & (0.058) \\
Automatic Transmission & -0.183*** & (0.013) \\
\midrule
% --- Engine Type Effects ---
4 Cylinder Engine & 0.396*** & (0.045) \\
6 Cylinder Engine & 0.441*** & (0.046) \\
8 Cylinder Engine & 0.686*** & (0.047) \\
10 Cylinder Engine & 0.789*** & (0.085) \\
12 Cylinder Engine & 0.685*** & (0.056) \\
\midrule
% --- Year Effects ---
Year 2015 & 0.172 & (0.442) \\
Year 2016 & 0.255 & (0.424) \\
Year 2017 & 0.251 & (0.422) \\
Year 2018 & 0.390 & (0.421) \\
Year 2019 & 0.364 & (0.421) \\
Year 2020 & 0.512 & (0.421) \\
Year 2021 & 0.721 & (0.422) \\
Year 2022 & 0.762 & (0.423) \\
Year 2023 & 0.674 & (0.425) \\
\bottomrule
\end{tabular}

\begin{tablenotes}
    \item[\small] \textbf{Note:} Dependent variable is log(winning\_bid). Model estimated using pooled OLS with cohort and time dummies. Standard errors in parentheses.
    
    \item[\small] \textbf{Model Statistics:} R² = 0.5901, N = 17,695, RMSE = 0.5905, F-stat = 129.85
    
    \item[\small] \textbf{Significance:} *** p < 0.001, ** p < 0.01, * p < 0.05
    
    \item[\small] \textbf{Variable Descriptions:}
    \begin{itemize} \setlength\itemsep{0.5pt}
        \item Log(Mileage): Natural log of vehicle mileage + 1
        \item Log(Horsepower): Natural log of vehicle horsepower  
        \item Vehicle Age: Age of vehicle at time of sale
        \item Time Trend: Normalized time trend (0 to 1)
        \item Automatic Transmission: 1 if automatic, 0 if manual
        \item Engine dummies: Reference category is 2 Cylinder
        \item Year effects: Annual time fixed effects
    \end{itemize}
\end{tablenotes}

\end{threeparttable}
\end{table}

Table~\ref{tab:hedonic_parameters} presents the coefficient estimates from the hedonic model. The model explains approximately 59\% of the variation in log price in the training sample ($R^2 = 0.5901$), with a root mean squared error (RMSE) of 0.5905 log points. The coefficient on log mileage implies that a 10\% increase in mileage is associated with a 0.77\% decrease in auction price, \emph{ceteris paribus}. This elasticity is consistent with buyer preferences for lower-use vehicles and suggests that mileage-based depreciation is a first-order determinant of consumer valuation. Similarly, the coefficient on log horsepower indicates that a 10\% increase in horsepower raises price by approximately 2.8\%, \emph{ceteris paribus}. These estimates confirm that mechanical freshness and performance attributes remain salient in the used enthusiast vehicle market.

We estimate a statistically significant negative premium for automatic transmissions: vehicles with automatic gearboxes sell for 18.3\% less than their manual counterparts, \emph{ceteris paribus}. This is consistent with the platform’s positioning toward driving enthusiasts, many of whom view manual transmissions as more desirable. The result also illustrates the value of platform-specific preference estimates: unlike mass-market pricing, enthusiast markets may penalize automatic variants of otherwise similar vehicles. The coefficient on vehicle age is $-0.001$ (p~<~0.01), which corresponds to a modest 0.1\% decline in price per additional calendar year. 

We observe a monotonic and economically large premium associated with higher cylinder counts. Compared to the omitted 2-cylinder category, 4-cylinder engines sell for 39.6\% more, 6-cylinder engines for 44.1\% more, and 8-cylinder engines for 68.6\% more (all p~<~0.001). The premium peaks for 10-cylinder engines (+78.9\%), while 12-cylinder configurations also command a large premium (+68.5\%). These results suggest that cylinder count captures not only horsepower but also brand prestige, perceived performance potential, and rarity. The especially high valuations for 10- and 12-cylinder configurations reflect their association with high-end or exotic models, even after controlling for brand via cohort effects.

The time trend variable is positive ($0.102$) but not statistically significant at conventional thresholds, suggesting that linear trends are largely absorbed by the year dummies. Year fixed effects show a marked increase in transaction prices from 2015 onward, peaking in 2022 with a coefficient of $0.762$. This implies that, relative to the base year (2014), average auction prices more than doubled by 2022. The pattern aligns with widely documented pandemic-era price spikes in the used vehicle market, likely driven by supply chain disruptions, increased disposable income, and shifts in consumer behavior.

\section{Comparative Results of Two-Stage vs. Direct Estimation} \label{app:two-stage-v-direct}

To test the performance of our proposed model, we compare the results we achieve with our two-stage estimation procedure to that of a ``direct'' estimation procedure that bypasses this initial embedding step and instead trains the language model and the structural model simultaneously. For reference, Figure~\ref{fig:proposed-model} depicts the two-stage estimation procedure proposed in this paper. In the first stage, the language model $\mathcal{M}$ converts unstructured textual data into embeddings $\vec e$, which subsequently feed into an auxiliary decoding model $\mathcal{H}_1$. The model $\mathcal{H}_1$ generates preliminary estimates ($\hat{Z}$), compared against known intermediate outcomes ($Z$) to minimize a first-stage loss function ($\mathcal{L}_1$). This step explicitly trains embeddings to represent economically relevant information, such as product quality signals or market conditions. In the second stage, these embeddings are input into a structural econometric model ($\mathcal{H}_2$), which estimates bidder valuations and market outcomes ($\hat{Y}$) by minimizing a second-stage loss function ($\mathcal{L}_2$) relative to observed targets ($Y$).

In contrast, the direct estimation procedure bypasses the first-stage decoding model ($\mathcal{H}_1$) entirely, jointly training the language and structural models using a single loss function. This approach directly optimizes the embeddings and structural parameters simultaneously, without explicit intermediate supervision guiding the embedding representation.

\begin{figure}
    \centering
    \includegraphics[width=0.75\linewidth]{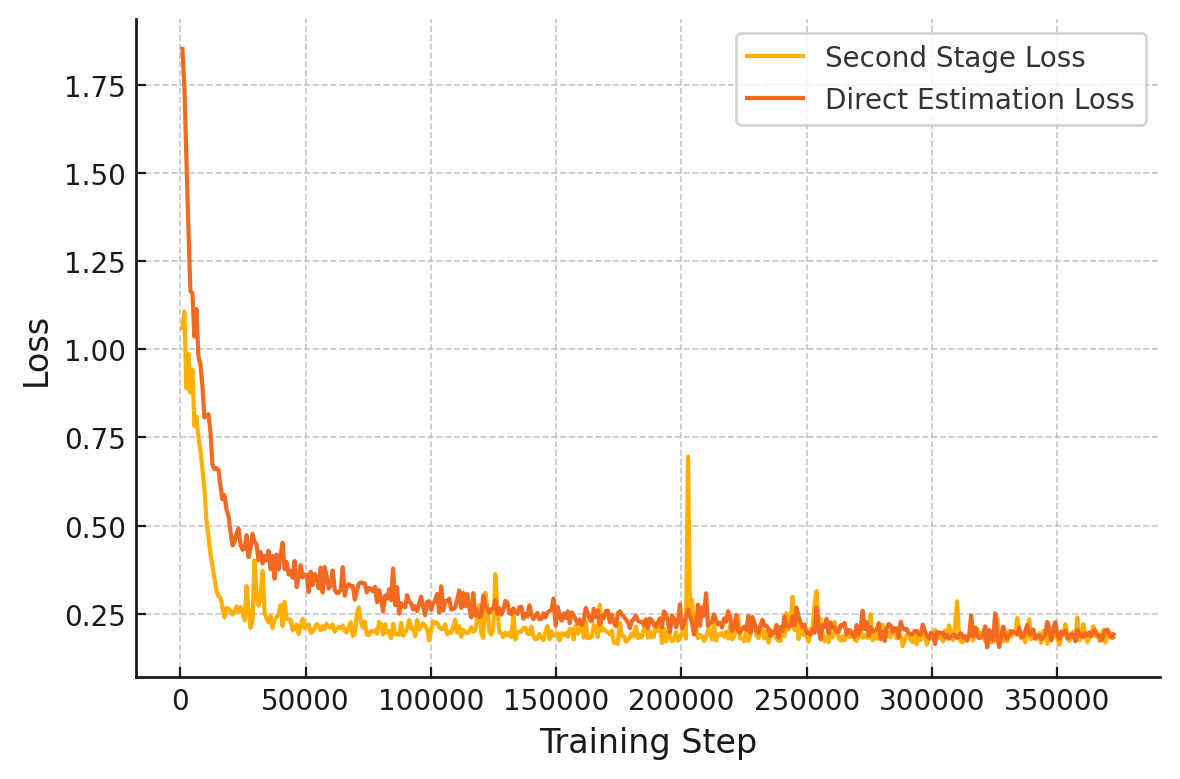}
    \caption{Training Results (Loss) for Two-Stage vs. Direct Estimation}
    \label{fig:training-loss-comparison}
\end{figure}

To evaluate differences in empirical performance, we compared the training loss trajectories of both approaches in Figure \ref{fig:training-loss-comparison}. The loss curves exhibit strong correlation throughout training (Pearson $r = 0.91$), and examining convergence performance over the final 5,000 training steps, we observe that both methods reach statistically indistinguishable levels of loss (mean absolute difference = $0.015$; paired $t$-test, $p=0.70$). These findings imply that while the two-stage procedure provides an advantage in early-stage learning efficiency, both estimation methods ultimately achieve comparable levels of fit to the training data.

\begin{figure}
    \centering
    \includegraphics[width=0.85\linewidth]{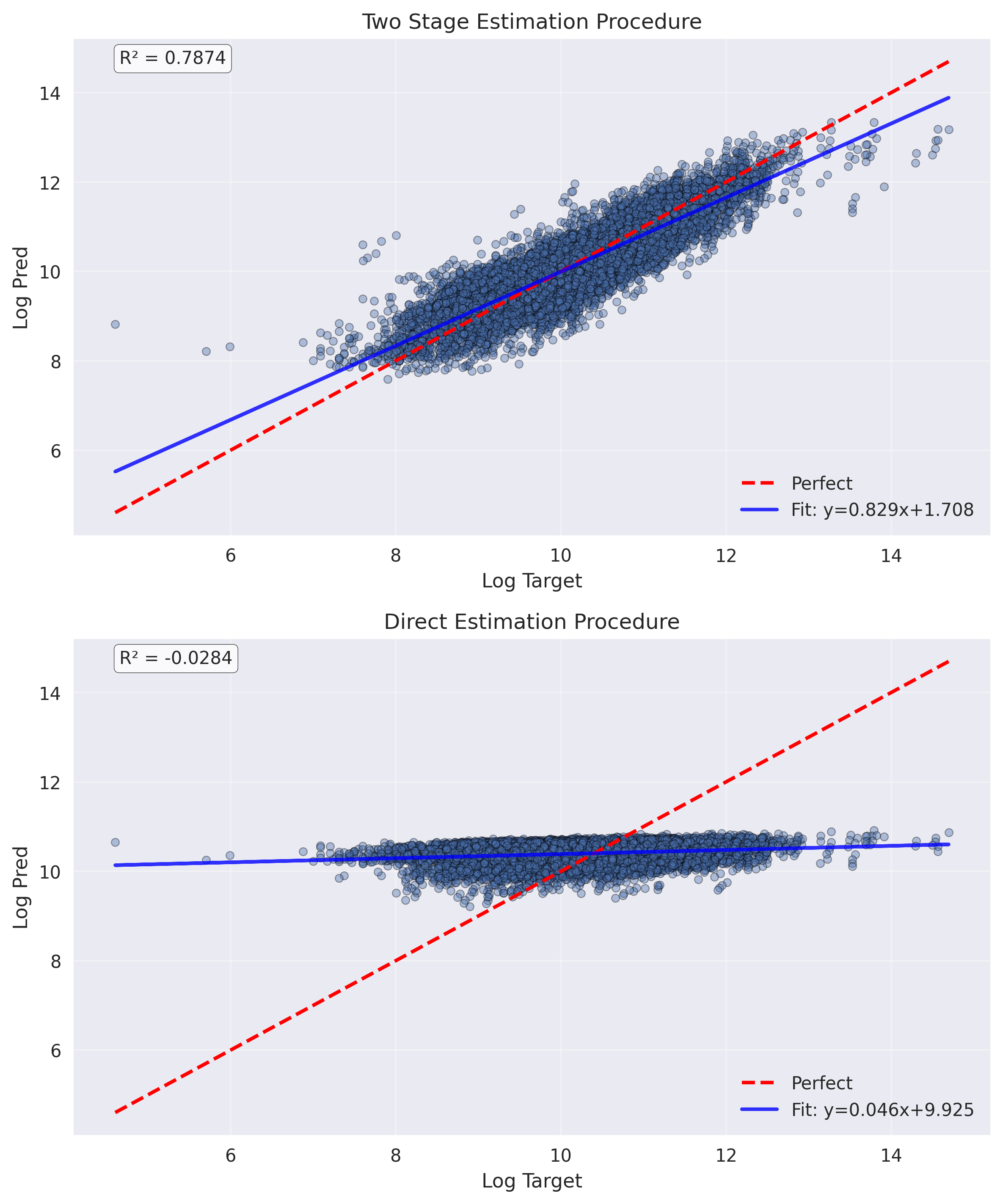}
    \caption{Validation Results for Two Stage vs. Direct Estimation}
    \label{fig:scatter-comparison-log}
\end{figure}

However, on our holdout validation set comprised of all 2$^\text{nd}$- to 5$^\text{th}$-bids aggregated from 5,000 auctions, the direct estimation procedure exhibits virtually no predictive power. In Figure \ref{fig:scatter-comparison-log}, the fitted relationship between log‐predictions and log‐targets is nearly flat (slope of 0.046) and yields a negative out‑of‑sample $R^2$ of –0.028, indicating performance worse than a constant‐mean benchmark. In practice, predicted log prices collapse around the sample mean ($\approx$10), capturing almost none of the true variation in bidder behavior. By contrast, the two‑stage estimator achieves an out‑of‑sample $R^2$ of 0.79 with a slope of 0.83, demonstrating substantially superior generalization. These results highlight that, absent the explicit embedding supervision of Stage 1, joint end‑to‑end training collapses to trivial, uninformative predictions.  

\section{Additional Graphical Analyses} \label{app:additional-analyses}

\begin{figure}
\centering 
\begin{tikzpicture}
\begin{axis}[
    hide axis,
    scale only axis,
    height=0pt,
    width=10cm,
    colorbar horizontal,
    point meta min=-1,              % scores range –1 … +1
    point meta max= 1,
    colorbar style={
        width=10cm,
        height=0.5cm,
        xtick={-1,-0.5,0,0.5,1},
        xticklabels={$-1$,$-0.5$,$0$,$0.5$,$1$},
        xticklabel style={text height=1.5ex},
        xtick style={draw=none},
        colormap name=divRedWhiteBlue
    },
]
    % dummy invisible plot just to force the bar to draw
    \addplot [draw=none] coordinates {(-1,0) (1,0)};
\end{axis}
\end{tikzpicture}

\raggedright 
\input{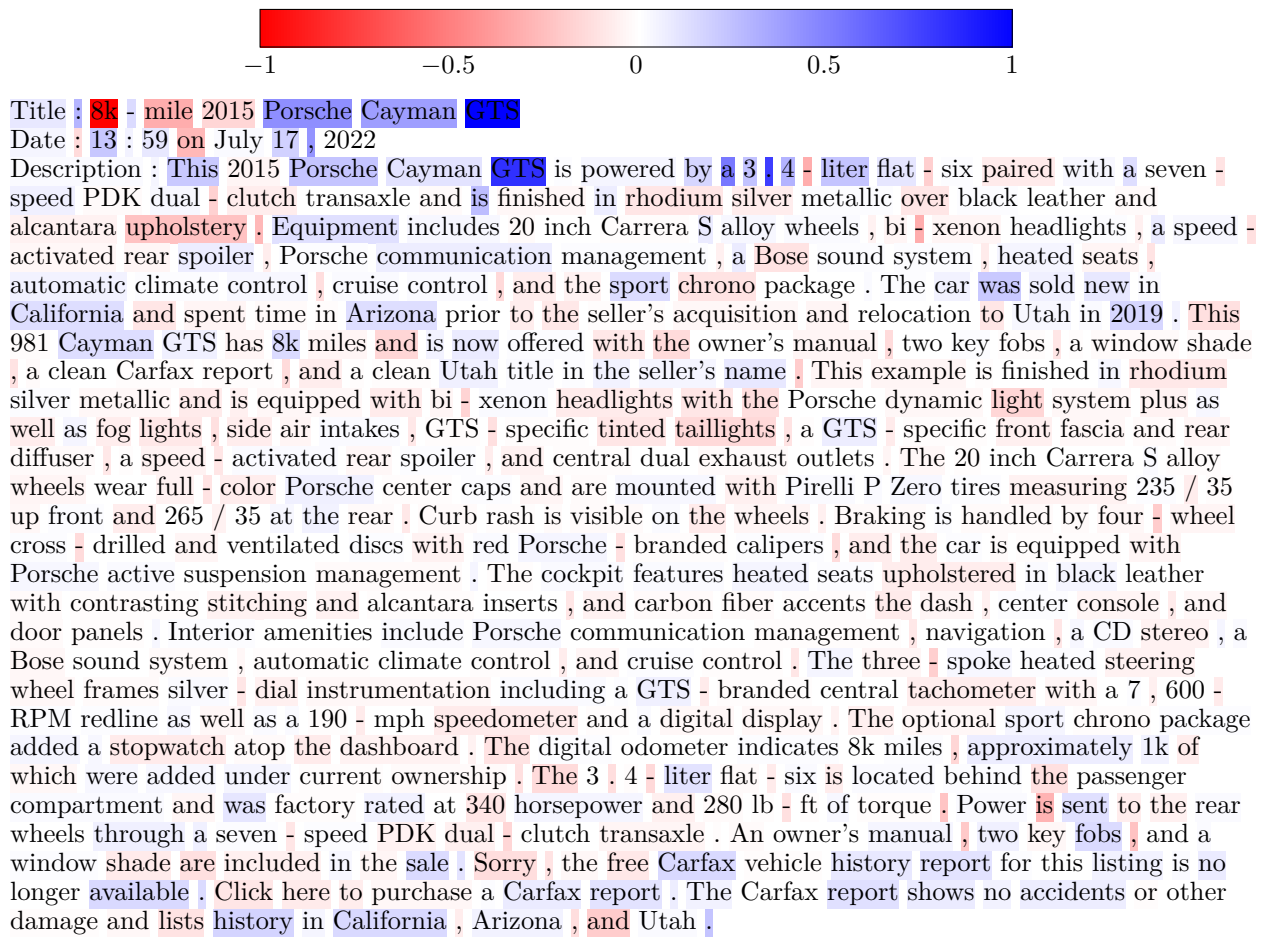}
\caption{Stage 1 Estimation: Normalized Token Contributions to Number of Auction Views}
\label{fig:stage-1-views-example}
\end{figure}

\end{document}